\documentclass{article}
\usepackage{enumerate}
\usepackage{natbib}
\usepackage{url} % not crucial - just used below for the URL 
\usepackage{listings}
\usepackage{numprint}
\usepackage{bbm}
\usepackage{graphicx,amscd,amsthm,amsmath,amssymb,amsfonts,verbatim,bm}
\usepackage{setspace}
\usepackage{fancyhdr}
\usepackage{titling}
\usepackage{authblk}
\usepackage{url}
\usepackage{enumitem}
\usepackage{multirow}
\usepackage[all]{xy}
\usepackage[flushleft]{threeparttable}
\usepackage[plain,noend]{algorithm2e}
\usepackage{amsmath}
\usepackage{scalerel}
\usepackage{adjustbox}
\usepackage{blindtext, rotating}
\usepackage{amsthm}
\usepackage{bbm}
\usepackage{amsmath}
\usepackage{amssymb}
\usepackage{natbib}
\usepackage{enumerate}
\usepackage{graphicx}
\usepackage{amssymb}
\usepackage{multirow}
\usepackage{verbatim}
\usepackage{subfigure,placeins}
\usepackage{array}
\newcolumntype{C}[1]{>{\centering\arraybackslash}p{#1}}

\usepackage{enumerate}
\usepackage{bbold}

\newcommand\R{\mathbb{R}}
\newcommand\E{\mathbb{E}}

\usepackage{bbm}
\usepackage{tikz}
\usepackage{multirow}
\usetikzlibrary{shapes,decorations,arrows,calc,arrows.meta,fit,automata,positioning}
\usetikzlibrary{shapes.misc}
\usetikzlibrary{shapes.geometric}
\usepackage{comment}
\usepackage{pgfplots}

\usepackage{xcolor}
\tikzset{
	-Latex,auto,node distance =1 cm and 1 cm,semithick,
	state/.style ={circle, draw, minimum width = 0.7 cm},
	box/.style ={rectangle, draw, minimum width = 0.7 cm, fill=lightgray},
	point/.style = {circle, draw, inner sep=0.08cm,fill,node contents={}},
	bidirected/.style={Latex-Latex,dashed},
	el/.style = {inner sep=3pt, align=left, sloped}
	
}
\usepackage[plain,noend]{algorithm2e}

\newtheorem{theorem}{Theorem}
\newtheorem{corollary}{Corollary}
\newtheorem{condition}{Condition}
\newtheorem{assumption}{Assumption}
\newtheorem{definition}{Definition}
\def\bV{\mathbf{V}}
\def\bU{\mathbf{U}}
\def\E{\mathbb{E}}
\newcommand{\norm}[1]{\left\lVert#1\right\rVert}

\def\bV{\mathbf{V}}
\def\E{\mathbb{E}}
\def\bU{\mathbf{U}}
\def\xistar{\xi^\ast}
\renewcommand\hat{\widehat}

\def\Ftrue{F^\ast}
\def\ftrue{f^\ast}
\def\ptrue{p^\ast}
\def\pitrue{\pi^\ast}
\def\thetatrue{\theta^\ast}
	\def\thetastar{\theta^\ast}
\title{Non-parametric Bayesian inference via loss functions under model misspecification}
	\author{Yu Luo\thanks{
		Department of Mathematics, King's College London, United Kingdom} ,\hspace{.2cm}
	David A. Stephens\thanks{
		Department of Mathematics and Statistics, McGill University, Canada}, \hspace{.2cm}
	Daniel J. Graham \thanks{
		Department of Civil and Environmental Engineering, Imperial College London, United Kingdom},\hspace{.2cm}
	Emma J. McCoy\thanks{
		Department of Statistics, London School of Economics and Political Science, United Kingdom}
}

\date{ }
	
\begin{document}
		
\maketitle

		\begin{abstract}
 In the usual Bayesian setting, a full probabilistic model is required to link the data and parameters, and the form of this model and the inference and prediction mechanisms are specified via de Finetti's representation.   In general, such a formulation is not robust to model misspecification of its component parts.  An alternative approach is to draw inference based on loss functions, where the quantity of interest is defined as a minimizer of some expected loss, and to construct posterior distributions based on the loss-based formulation; this strategy underpins the construction of the Gibbs posterior.  We develop a Bayesian non-parametric approach; specifically, we generalize the Bayesian bootstrap, and specify a Dirichlet process model for the distribution of the observables.  We implement this using direct prior-to-posterior calculations, but also using predictive sampling. We also study the assessment of posterior validity for non-standard Bayesian calculations.  We show that the developed non-standard Bayesian updating procedures yield valid posterior distributions in terms of consistency and asymptotic normality under model misspecification.  Simulation studies show that the proposed methods can recover the true value of the parameter under misspecification.

\noindent \textit{Key words:} General Bayesian updating; Bayesian loss learning; Model misspecification; Bayesian Non-parametric inference; Causal inference

\end{abstract}

\section{Introduction}
\label{sec:intro}

Bayesian inference methods are central to decision making under uncertainty.  The most common approach to Bayesian (prior-to-posterior) updating employs parametric specifications of probability models for the observable quantities, but there has also been much research on relaxing parametric assumptions, as parametric models are not typically robust to model misspecification; that is, they rely on correct specification of the terms that appear in de Finetti's representation.  In standard prior-to-posterior inference, coherent Bayesian updating of prior beliefs on the parameter follows from an assumption of exchangeability of the observable quantities, the de Finetti representation for the corresponding probability model, and the combination of prior distribution on the unobservable data generating model with an induced conditional probability model for the observables. The standard Bayesian approach is generative, in the sense that it relies on a probability model for the observable quantities.  In contrast, \cite{zhang2006a} and \cite{bissiri2016general} formulate posterior inference entirely on a loss (or utility) specification to target a specific parameter in the data-generating distribution, leading to the so-called \textit{Gibbs posterior}.  The Gibbs posterior results from a prior-to-posterior update where the loss function is converted to yield a pseudo-likelihood, and then combined with a prior distribution.  An advantage of the targeting of a specific parameter of interest is that a full probabilistic specification for the data generating model for the observables is avoided.  However, in general a probabilistic interpretation of the assumptions concerning the distribution of the observables is lost. Another drawback of the Gibbs posterior is its difficulty in handling multi-parameter problems, as it may fail to provide valid marginal inference for individual parameters without additional calibration. An alternative construction adopts a predictive viewpoint.  \cite{fong2021martingale} proposed a \textit{martingale posterior}, calibrating the source of statistical uncertainty in Bayesian inference as the missing information from the as yet unobserved data in the observation sequence.  They proposed a copula-based predictive density to calculate the posterior distribution. Under exchangeability, the predictive approach is essentially equivalent to the conventional Bayes approach, but one which gives the potential for more general model constructions.  However, the martingale posterior approach offers the possibility of Bayesian modelling beyond exchangeability.

A common element in these approaches is that the quantities of interest are expressed as some functional of the data generating process, which is characterized by the distribution $\Ftrue$.  In any inference problem, $\Ftrue$ --- and thus any functional of it --- is regarded as an unknown quantity, and uncertainty is present due to absence of perfect knowledge of $\Ftrue$.  If prior uncertainty for $\Ftrue$ is encapsulated in a prior distribution, the form of the posterior on $\Ftrue$, and the posterior of any parameter of interest, can be deduced.  In the standard Bayesian approach, a `parameter' is defined as a functional of the limiting distribution of the exchangeable observables.  In a parametric specification, $\Ftrue(z) \equiv \Ftrue(z|\xistar)$, where $\xistar$ lies in the finite dimensional parameter space $\Xi$.  In the simplest case of exchangeable binary variables $\{Z_i,i=1,\ldots,n,\ldots\}$, for any $n \geq 1$, we have that for any binary sequence $z_1,\ldots,z_n$, the joint distribution can be written using the de Finetti representation as the integral over quantity $\xi$ of the product of the conditional densities $\ftrue(z_i|\xi) = \xi^{z_i} (1-\xi)^{1-{z_i}}$ and prior $\pi_0(\xi)$, where the true (data generating) parameter and distribution are
\[
\xi^\ast = \lim_{n \longrightarrow \infty} \frac{1}{n} \sum_{i=1}^n Z_i \quad
\text{and} \quad
\Ftrue((-\infty,z]) = \lim_{n \longrightarrow \infty} \frac{1}{n} \sum_{i=1}^n \mathbb{1}_{(-\infty,z]}(Z_i) \quad z \in \R
\]
respectively.  The de Finetti representation defines `parameters' in this way, although alternative definitions can also be used -- see \citet[Chap.4]{bernardo2009bayesian}. A formulation that identifies the same parameter is based on loss minimization, with
\[
\xi^\ast = \arg \min_t \int - \log \ftrue(z|t) d \Ftrue(z) = \arg \max_t \int (z \log t + (1-z) \log (1-t))  d \Ftrue(z).
\]
The loss function $\ell(z,t) = - \log \ftrue(z|t)$ defines the parameter as that which minimizes the expected loss under $\Ftrue (z|\xi^\ast)$.  Note that we may equivalently define $\xi^\ast$ via a different loss function and functional of $\Ftrue$: for example, using $\ell(z,t) = \lambda (z-t)^2$ with parameter $\lambda > 0$ returns the same minimizer, so the formulation via a loss function minimization is not unique, and other aspects of inference such as uncertainty intervals in general depend on the loss function, illustrating the importance of `modelling' the connection between observables and the parameter.

The approach proposed in this paper is based entirely on Bayesian decision theory and the assumption of exchangeability, but could be extended to any generative modelling setting.   We conduct Bayesian inference, first via the usual prior-to-posterior procedure, and then via a predictive-to-posterior procedure. In particular, this work makes three specific contributions:
\begin{itemize}[nosep]
	\item We bridge traditional Bayesian updating and decision-theoretic predictive inference by formulating scalable posterior inference under non-parametric Dirichlet process (DP) representations. We propose two computationally efficient algorithms that construct posterior distributions via optimization rather than MCMC. Asymptotically, we demonstrate that the two updating frameworks yield the same posterior distribution under the exchangeability assumption and guarantee consistent estimation under mild conditions. 
	\item We establish the validity of the resulting posterior inference in terms of uncertainty quantification, extending the notion of proper posteriors in the spirit of \cite{monahan1992proper} to non-standard Bayesian updating schemes. 
	\item We develop a variance-matching calibration approach for the learning rate in the Gibbs posterior with the posterior obtained from the proposed non-parametric inference algorithms.
\end{itemize}

Beyond these methodological advances, this work contributes to the Bayesian causal inference literature by enabling efficient and robust causal inference under model misspecification. The causal inference example is selected to demonstrate the theoretical and empirical usefulness of our proposed Bayesian loss function approach, which involves the use of propensity score (PS) adjustment in doubly robust (DR) causal inference.  DR procedures have a well-established basis in frequentist semi-parametric theory, with estimation of causal parameters typically conducted via outcome regression (OR) and PS adjustment. The key feature of DR models is that consistent estimation of a typical causal estimand, the average treatment effect (ATE), requires only one of the OR or PS models to be correctly specified, thus adding a degree of robustness in the estimation of causal quantities. A Bayesian approach to semi-parametric DR inference is not obvious as it typically avoids specification of a likelihood function.  Most of the proposed methods employ two-stage PS adjustment or flexible outcome modeling \citep[see][]{stephens2021bayesian}, with several recent work on Bayesian semi-parametric methods \citep[for example,][]{graham2016approximate, saarela2016bayesian,luo2022semi}; these methods typically exploit computational approaches, in particular the Bayesian bootstrap to perform inference.  However, to our knowledge, there has not been work on such Bayesian causal inference procedures  via the proposed loss updating under the Bayesian non-parametric regime nor assessing the validity of such Bayesian causal procedures. In Section \ref{BayesLoss}, we formulate the assessment of posterior validity on loss-based inference. We introduce the Bayesian non-parametric modelling in two different updating framework in Section \ref{Sec:postloss}.  Section \ref{sec:AP} outlines the asymptotic justification for the proposed approach, followed by the example of Bayesian DR causal inference via an augmented OR in Section \ref{bayecausal}. Section \ref{Sec:sim} demonstrates the proposed method with simulation studies and a real-data example, following with concluding remarks and future research directions in Section \ref{Sec:dis}.

\section{Validity of Bayesian inference via loss functions}
\label{BayesLoss}

\subsection{Correct specification and misspecification}
For standard Bayesian inference under exchangeability, a full probabilistic model is required for the joint distribution of observable quantities, $p(z_1,\ldots,z_n), n \geq 1$.   The parametric case is formulated via a probability model $f(z|\xi)$ for conditionally independent and identically distributed observable random variables $Z_i, i=1,\ldots,n$, and focuses on the posterior distribution, $\pi\left(\xi\left|z\right.\right) \propto \mathcal{L}\left(\xi\right) \pi_0\left(\xi\right)$, where $\mathcal{L}\left(\xi\right)=\prod_{i=1}^{n} f\left(z_i\left|\xi\right.\right)$ is the likelihood and $\pi_0\left(\xi\right) $ the prior density for $\xi$.  Most commonly, inference proceeds under the assumption of `correct model specification'.  This concept is more nuanced than in the frequentist sequel. A strict interpretation would require that $f(z|\cdot)$ is the correct (data generating) probability model, $\xi^\ast$ is the true data generating parameter, and $Z_1,\ldots,Z_n$ are independently drawn from $f(z|\xi^\ast)$.  This may be relaxed to require that the posterior $\pi\left(\xi\left|z\right.\right)$ converges to a point mass at $\xi^\ast$, so that the data generating model is recovered in the limit; this effectively requires that $\xi^\ast$ lies in the support of $ \pi_0$, although the requirement that $f(z|\cdot)$ is the correct conditional model is retained.  The usual approach to misspecification proceeds assuming that the conditional model does not coincide with the data generating conditional model.  Henceforth, we denote the correct conditional model, generic parameter, and parameter true value by $\ftrue$, $\xi$ and $\xi^\ast$, and a competing model by $f$ parameterized by $\theta \in \Theta$.  

\subsection{{Targeting parameters via loss minimization}}

Under exchangeability, $\xi^\ast$ is well-defined as a limiting random variable that is a function of the observables.  The `true value' of the parameter, $\theta^\ast$, in the misspecified model may be identical to $\xi^\ast$, or an element or subvector of $\xi^\ast$, or a parameter defined functionally via $\Ftrue$.  We proceed by assuming that $\theta^\ast$ is defined as the value which minimizes an expected loss
\begin{equation}
	\label{u2}
	\theta^\ast=\arg\min\limits_{\theta \in \Theta }\mathbb{E}\left[\ell\left(Z,\theta\right)\right]=\arg\min\limits_{\theta \in \Theta}\int \ell\left(z,\theta\right) d\Ftrue\left(z\right)
\end{equation}
where the integral is presumed finite for at least one $\theta \in \Theta$.   Under the assumption that $\Ftrue$ admits a parametric density, then $\theta^\ast=\arg\min\limits_{\theta \in \Theta}\int \ell\left(z,\theta\right) \ \ftrue(z|\xi^\ast) \ dz$
and it is evident that inference about $\theta$ follows from inference about $\xi$. 

The loss function $\ell(z,\theta)$ captures the loss in the proposed alternative inference model.  In the parametric case, if $ \ell (z,t ) = -\log f  (z |t  )$ for some other density $f$, the expectation becomes (up to an additive constant that does not depend on $t$) the Kullback-Leibler (KL) divergence between the true model $\ftrue (z | \xistar )$ and $f(z | t )$.  This case remains within a probabilistic model framework. However, loss-based inference also encompasses situations where no statistical model is assumed, making it irrelevant to question whether the model is (partially) misspecified, as seen in cases such as classification and quantile inference, which is commonly used in machine learning. Therefore, we define misspecification in the context of loss function as follows:
\begin{definition}
	Consider a statistical model $\mathcal{F} = \left\{\ftrue(.|\xi): \xi  \in \Xi\right\}$, where $\Xi$ may be finite- or infinite-dimensional. Suppose $(Z_1,\ldots,Z_n)$ are i.i.d from $\ftrue(.|\xi^\ast)$. Consider  a loss function $ \ell(z,\cdot): \Theta \rightarrow \mathbb{R}$, such that for every $z$ in the support of the true distribution, the map $\ell(z,\cdot)$ is measureable with respect to $\ftrue(.|\xi^\ast)$. We say that the loss function is misspecified model if and only if $\ell (z,t ) \ne  -\log \ftrue(z | t)$ for some for some $(z,t)\in \mathcal{Z} \times \Theta$.  
\end{definition} 
If the model is correctly specified, and identifiable, then $\theta^\ast= \xistar$. If $f$ is misspecified, this definition of the `true' parameter is in line with standard frequentist arguments. If we further assume $\ell\left(z,\theta\right)$ is differentiable with respect to $\theta \in \Theta$ for all $z$ and {$\theta^\ast$ is identifiable in $\Theta$, then $\theta^\ast$  is the unique solution of the unbiased estimating equation}
\begin{equation}
	\label{est}
	\mathbf{U}(\theta) =	\int \frac{\partial \ell\left(z,\theta\right)}{\partial \theta} d\Ftrue\left(z\right)= \mathbb{E}\left[\frac{\partial \ell\left(Z,\theta\right)}{\partial \theta} \right]=0.
\end{equation}
If $\Ftrue\left(z\right)$ is represented using a non-parametric specification, we can retain most of the parametric calculations but with $\theta^\ast$ as a functional of $\Ftrue$. This loss-based formulation allows for the possibility of the loss function $\ell(z,\theta)$ as a misspecified model.  If $\ell(z,\theta) \ne  -\log f  (z |\theta )$ then this posterior for $\theta$, however it is computed, will quantify the posterior uncertainty in a quantity that is connected to the data generating mechanism in an abstract way.  This posterior, in general, is of little practical use as it does not facilitate inference in a true quantity of interest, nor does it facilitate prediction.  The exception is when $\theta^\ast$ is a meaningful parameter in the data generating model -- in this case, computing the posterior is still a worthwhile pursuit.  This realization reinforces the notion that to guarantee this compatibility, the data-generating model must be represented using a non-parametric formulation. Before detailing our approach to loss-based posterior, we must confirm that computing a posterior via non-standard methods still produces valid probabilistic statements. In the next section, we follow \cite{monahan1992proper} to illustrate how this framework can assess the validity of such posterior inferences.

\subsection{Assessing the validity of non-standard posterior inference}
\label{sec:valid}

Consistent estimation of $\theta^\ast$ is a minimal requirement for any statistical procedure, but any posterior inference calculation method should also exhibit appropriate performance in a finite sample.  We adopt the approach introduced in \cite{monahan1992proper}, which addressed the notion of proper Bayesian inference when replacing the parametric likelihood with an alternative likelihood function (say, for example, a marginal or conditional likelihood). They argued that the `posterior' density, $\widetilde \pi(\theta|z_{1:n})$, computed by a non-standard method should still make probability statements consistent with the Bayes rule. 
\begin{definition}
	A `posterior' density, $\widetilde \pi(\theta|z_{1:n})$, is valid by coverage, for the data $z_1,\ldots,z_n \sim \Ftrue$, if its posterior credible sets, $\mathcal{C}_\kappa(z)$, achieve the nominal coverage under repeated sampling from $\Ftrue$.   
	That is,
	\[
	\int \left\{\int_{\mathcal{C}_\kappa(z)} \widetilde\pi(\theta|z_{1:n}) \ d \theta \right\} d \Ftrue(z) = 1- \kappa.
	\]
\end{definition}
This definition requires that a vaild posterior distribution, $\widetilde \pi(\theta|z_{1:n})$, yields coverage at the nominal level.	Assuming for the moment correct specification, $\mathcal{C}_\kappa(z)$ should achieve nominal coverage under data generating joint measure of $Z$ and $\theta$, that is, $P_{\widetilde \pi} (\theta \in \mathcal{C}_\kappa(z))$ should have expectation $1-\kappa$ for data generated under the measure  {$\pi_0(\theta) \ftrue(z|\theta)$} for every absolutely continuous prior, $\pi_0(\ldotp)$. We may assess this using the probability integral-transformed random variable
\begin{equation}\label{Hformula}
	H = \int_{-\infty} ^{\theta} \widetilde \pi(t|z_{1:n}) \ d t = P_{\widetilde \pi} (T \leq \theta).
\end{equation}
If $\widetilde \pi(\theta|z_{1:n})$ is a valid posterior, $H \sim Uniform(0,1)$. In simulation context, if we generate $\theta^{(k)} (k=1,\ldots,m) \sim \pi_0(\ldotp)$ and data, $z_{1:n}^{(k)}$, from  $\ftrue( \ldotp |\theta_k)$ (given the true data generating model). Based the data, we can compute the putative posterior $\widetilde \pi(\theta|z_{1:n}^{(k)})$. Then we can obtain $H_k$ based on \eqref{Hformula} by replacing $\theta$ with $\theta^{(k)}$.   If the distribution of $H_k$ follows the uniform distribution, then $\widetilde \pi(\theta|z_{1:n})$ is defined as a proper coverage, yielding valid posterior inference.  On the other hand, this method can be viewed as an assessment purely from the Bayesian decision theory. In  Bayesian decision theory, the optimal Bayes action, $a^\ast(z)$,  is found by maximizing expected utility or minimizing the posterior expected loss
	\[
	a^\ast(z) = \arg\min\limits_{a(z) \in \mathcal{A}} \int \ell(\theta, a(z))\frac{\pi_0(\theta) \ftrue(z|\theta)}{\displaystyle \int\pi_0(t) \ftrue(z|t) \ d t} \ d \theta. 
	\]
	If we take the quantile-based loss as $
	\ell(\theta, a(z))  = \kappa(a(z)-\theta )\mathbb{1}(\theta < a(z)) + (1-\kappa)(\theta - a(z))\mathbb{1}(\theta \ge a(z))$,
	then the optimal action is the $100 \times (1-\kappa)$\% quantile of the posterior distribution, i.e., $\P(\theta < a^\ast(z)|z_{1:n}) = 1-\kappa$.  Similar to the Monahan and Boos method, any valid posterior density, $\widetilde{\pi}(\theta|z_{1:n
	})$, should result in the same quantile given the optimal action, $a^\ast(z)$, i.e.,  $\mathbb{P}_{\widetilde \pi}(\theta < a^\ast (z)| z_{1:n})=1-\kappa$. 
	
	\vspace{-0.2in}
	
	\section{Posterior calculation via loss functions}
	\label{Sec:postloss}
	\subsection{Prior-to-posterior updating via Bayesian non-parametric modelling}
	In this section, we will layout our approach to calculate loss-based posterior from two different perspectives. Consider the right-hand side of equation \eqref{u2}, if we have a posterior distribution $\pitrue(\xi|z_{1:n})$ based on the true data generating model, $\Ftrue(z|\xi^\ast)$, then the uncertainty represented by the posterior is preserved under the deterministic calculation implied by \eqref{u2}; that is, for example if $\xi^{s}$ is a sampled variate from $\pitrue(\xi|z_{1:n})$, then the quantity
	\begin{equation}
		\label{u2samp}
		\theta^{s}=\arg\min\limits_{\theta \in \Theta}\int \ell\left(z,\theta\right) d \Ftrue(z|\xi^{s})
	\end{equation}
	is a variate drawn from the posterior for $\theta$. The parametric version of  calculation in  \eqref{u2samp} relies on correct specification of the model leading to the calculation of posterior $\pitrue(\xi|z_{1:n})$ to guarantee consistent estimation. A Bayesian non-parametric formulation gives protection against misspecification; we henceforth denote a generic instance $F$.   In this case, the posterior distribution for $F$ is a probability distribution on the space of distribution functions, and a draw from this posterior is a random distribution which can be transformed into a sampled variate $\theta$, which may be replicated to reproduce the posterior for the minimizing quantity as indicated by \eqref{u2}.

	A simple implementation of this Bayesian non-parametric theory is given by the Bayesian bootstrap \citep{rubin1981bayesian}, which assumes that the data points are realizations from a multinomial model on the values $\left(z_1,\ldots,z_n\right)$ with unknown probability $\varpi=\left(\varpi_1,\ldots,\varpi_n\right)$, and assumes a priori that $\varpi\sim \text{Dirichlet}\left(\alpha,\ldots,\alpha\right)$.  Then, a posteriori $\varpi\sim \text{Dirichlet}\left(\alpha+1,\ldots,\alpha+1\right)$. Conditional on a drawing $\varpi$ from the posterior distribution, posterior predictive samples can be drawn independently from $\{z_1,\ldots,z_n\}$ with associated probabilities $\{\varpi_1,\ldots,\varpi_n\}$.  The Bayesian bootstrap is obtained under the improper specification $\alpha=0$.  Referring to \eqref{u2samp}, the parameter $\theta$ can then be derived via 
	\begin{equation}\label{wlike}
		\theta  (\varpi )=\arg\min_{\theta\in \Theta}\sum_{k=1}^{n}\varpi_k\ell\left(z_k,\theta\right)
	\end{equation}
	that is, via a deterministic transformation of $\varpi$. A sample from the posterior distribution for $\theta$ can be obtained by repeatedly drawing $\varpi \sim \text{Dirichlet}\left(1,\ldots,1\right)$ and obtaining the solutions to \eqref{wlike}.  {The Bayesian bootstrap is a consequence of a DP specification that can be implemented in a more general form. Suppose that, a priori, $F \sim DP\left(\alpha,G_0\right)$ where $\alpha>0$ is the concentration parameter and $G_0$ is the base measure. In light of data $\left(z_1,\ldots,z_n\right)$, the resulting posterior distribution of $F$ is $DP\left(\alpha_n,G_n\right)$, where $\alpha_n=\alpha+n$ and $G_n(\cdotp)= \alpha G_0(\cdotp) \big /{(\alpha+n)} + {\sum_{k=1}^{n}\delta_{z_k}\left(\cdotp\right)}\big/{(\alpha+n)}$, and the posterior predictive distribution is effectively identical to the posterior distribution; a random draw from posterior distribution on $F$ provides a (conditional) distribution from which the observables may be drawn independently. If $\alpha \longrightarrow 0$, the posterior distribution is realized as a $\text{Dirichlet}\left(1,\ldots,1\right)$ distribution on $\left\{z_1,\ldots,z_n\right\}$, which reduces to the distribution implied in the Bayesian bootstrap.  If $\alpha >0$, this is still a standard DP model specification, but where $\left\{\zeta_k\right\}_{k=1}^{\infty} \sim G_n$ and $\left\{\varpi_k\right\}_{k=1}^{\infty} \sim StickBreaking(\alpha_n)$; the standard stick-breaking algorithm \citep{sethuraman1994constructive} generates the weights by a transformation of the collection $\{V_k\}_{k=1}^\infty$ where random variables $V_k \sim Beta(1,\alpha)$ are independent, with $\varpi_1 = V_1$ and for $j=2,3,\ldots$, $	\varpi_j = V_j \prod_{k=1}^{j-1} (1-V_k)$. 	In the case $\alpha > 0$, the equivalent to \eqref{wlike} is
	\begin{equation}\label{wlikeInf}
		\theta (\varpi,\zeta )=\arg\min_{\theta\in \Theta}\sum_{k=1}^{\infty}\varpi_k\ell\left(\zeta_k,\theta\right).
	\end{equation}
	Although this is an infinite sum, the $\varpi_k$ decreases in expectation as $k$ increases and eventually becomes numerically negligible (See the discussion in the Supplementary Material). The random draw of $\{\varpi_k,\zeta_k\}_{k=1}^\infty$ is then converted by the deterministic transform implicit in \eqref{wlikeInf} into a sample from the posterior for the target parameter $\theta$.  Algorithm \ref{A0} describes this approach.

	\begin{figure}[h]
		\centering

		\begin{minipage}[t]{0.48\textwidth}
			\begin{algorithm}[H]
				\small
				\renewcommand{\arraystretch}{0.5}
				\SetAlgoLined
				\KwData{$z_{1:n}=(z_1,\ldots,z_n)$}
				\KwIn{Given $G_0$, $\alpha$ and $N$.}
				\For{$s$ \textbf{to} $1:S$}{
					Sample $\{\zeta^{s}_k\}_{k=1}^N \sim G_{n}$ independently\;
					Sample $\{\varpi_k^{s}\}_{k=1}^N$ from a stick-breaking process with $\alpha_n=\alpha+n$\;
					Compute $\theta^{s}$ by solving the minimization problem in \eqref{wlikeInf}\;
				}
				\Return $(\theta^{1},\ldots,\theta^{S})$\;
				\caption{ \small \label{A0}Prior-to-posterior inference based on a stick-breaking process.}
			\end{algorithm}
		\end{minipage}
		\hfill
		\begin{minipage}[t]{0.48\textwidth}
			\begin{algorithm}[H]
				\small
				\renewcommand{\arraystretch}{0.5}
				\SetAlgoLined
				\KwData{$z_{1:n}=(z_1,\ldots,z_n)$}
				\KwIn{Given $G_0$, $\alpha$ and $N$.}
				\For{$s$ \textbf{to} $1:S$}{
					\For{$j$ \textbf{to} $1:N$}{
						Sample $z^{s}_j \sim G_{n+j-1}$\;
						Update $G_{n+j} \leftarrow \{z^{s}_j,G_{n+j-1}\}$\;
					}
					Obtain $z^{s}=\{z^{s}_{1},\ldots,z^{s}_{N}\}$\;
					Compute $\theta^{s}$ by solving the minimization problem in \eqref{MCminimizer}\;
				}
				\Return $(\theta^{1},\ldots,\theta^{S})$\;
				\caption{\small\label{A1}Predictive-to-posterior inference based on a P\'olya urn scheme.}
			\end{algorithm}
		\end{minipage}
		
	\end{figure}

	\subsection{Predictive-to-posterior updating via Bayesian non-parametric modelling}
	\label{predpost}
	In Bayesian decision theory, the Bayes estimator of a target parameter minimizes the posterior expected loss, given by $\hat \theta = \arg \min\limits_{t \in \Theta} \int_{\Xi} u\left(t,\xi \right) \pitrue\left(\xi | z_{1:n} \right) \ d \xi$ where $u$ is a real-valued function quantifying the loss between the $\theta$ and  $\xi$.  If the function $u$ is taken to be the KL divergence,
	\[
	u\left(\theta, \xi\right)= \int  \log\left(\dfrac{\ftrue(z | \xi )}{f(z | \theta )}\right) \ftrue(z | \xi )  \ dz
	\]
	then the minimizing value of $\theta$ is that for which
	$\int \log f (z |\theta) \left\{ \int_{\Xi} \ftrue(z | \xi) \pitrue(\xi | z_{1:n} )  \ d \xi \right\} \ dz \equiv
	\int \log f (z | \theta)  \ \ptrue (z|z_{1:n}) \ d z$
	is maximized, where $\ptrue(z|z_{1:n})$ is the usual Bayesian posterior predictive distribution.  Consider a new set of exchangeable data, $z^s_{1},\ldots,z^s_{N}$, and take $u\left(\theta,\xi \right)$ as
	\[
	\mathbb{E}_{Z^s_{1:N}\mid \xi}\left[\ell\left(Z^s_{1:N},\theta\right)\right]= \sum_{i=1}^{N}\mathbb{E}_{Z^s_{i}\left|\xi\right.}\left[\ell\left(Z^s_{i},\theta\right)\right] = \sum_{i=1}^{N} \int \ell (z^s_{i},\theta ) \ftrue(z_i^s \mid \xi) \ dz_i^s,
	\]
	that is, the expected loss under the `correct specification' that presumes {$ F= \Ftrue$}.  Then
	\begin{equation}
		\label{pp}
			\arg \min\limits_{\theta \in \Theta} \int_{\Xi} \sum_{i=1}^{N}\mathbb{E}_{Z^s_{i}\left|\xi\right.}\left[\ell\left(Z^s_{i},\theta\right)\right] \pitrue\left(\xi \left|z_{1:n}\right.\right)d\xi =\arg \min\limits_{\theta \in \Theta} \int \sum_{i=1}^{N}\ell\left(z^s_i,\theta\right) \ptrue (z^s_{1},\ldots,z^s_{N} | z_{1:n} ) \ dz^s 
	\end{equation}
	where $\ptrue(z^s_{1},\ldots,z^s_{N} | z_{1:n} )\equiv\ptrue(z^s_{1:N}| z_{1:n} )$ is the $N$-fold posterior predictive distribution. Therefore, the solution to the minimization problem \eqref{pp} is the Bayesian estimator that mimics the calculation in \eqref{u2}, with the posterior predictive distribution replacing $\Ftrue\left(z\left| \xistar\right.\right)$. Therefore, we aim to find a predictive distribution as our best Bayesian estimate for {$\Ftrue\left(z\left| \xistar\right.\right)$}.

	{The integral in \eqref{pp} may not be analytically tractable, but typically may be approximated using Monte Carlo methods.  Conditioning on $\ptrue(z^s_{1:N}| z_{1:n} )$, if $z^{M}=\left(z^M_{1},\ldots,z^M_{N}\right)$ represent one sample drawn from  this predictive distribution, then the finite sample approximation to \eqref{pp} is
		\begin{equation}\label{MCminimizer}
			\theta\left(z^M\right) =\arg\min_{t}  \sum_{i=1}^{N}\ell\left(z^{M}_i,t\right).
		\end{equation}
		As $N\longrightarrow \infty$, under mild regularity conditions, the minimizer from \eqref{MCminimizer} converges to $\theta^\ast$ defined in \eqref{u2}. Following \citet{bernardo1979reference,bernardo2009bayesian} on the representation theorem under sufficiency, we have
		\begin{equation*}
			\begin{aligned}
				\ptrue(z^s_{1},\ldots,z^s_{N} | z_{1:n} ) &= \ptrue (z^s_{1},\ldots,z^s_{N} \mid \hat \xi (z_{1:n} ) ) + \textrm{O}(1)\\
				&= \ptrue(z^s_{N}  \mid z^s_{1},\ldots,z^s_{N-1},\xistar)\cdots   \ptrue( z^s_{1}  \mid \xistar )+ \textrm{o}(1) &  n\longrightarrow \infty\\
				&=\ftrue(z^s_{N}  \mid \xistar)\cdots \ftrue( z^s_{1}| \xistar)+ \textrm{o}(1)
			\end{aligned}
		\end{equation*}
		where $\hat \xi (z_{1:n} )$ is the estimate for $\xistar$ via the sufficient statistics using the observed data $z_{1:n}$.  Therefore,  a draw from the predictive $\ptrue(z^s_{1},\ldots,z^s_{N} | z_{1:n} )$ suitably simulates a collection of $N$ sample points from the true data generating model $\Ftrue(z|\xistar)$ as $n\longrightarrow \infty$.  The minimizer of \eqref{MCminimizer} will become degenerate at $\theta^\ast$ as both $N\longrightarrow \infty$ and $n\longrightarrow \infty$. }
	
	Now we consider $\ptrue(z^s_{1:N}| z_{1:n} )$ as a random distribution. For example, we can specify the $\ptrue(z^s_{1:N}|z_{1:n} )$ as a DP.	The draws can be generated directly via stick break or P{\'o}lya urn schemes \citep{blackwell1973ferguson} that integrate the posterior distribution and allow direct draws of variables from the predictive distribution in a dependent fashion.  We simulate $S$ datasets of size $N$, with each dataset $z^{s}=\{z^{s}_{1},\ldots,z^{s}_{N}\},s=1,\ldots S,$ where $z^{s}$ is generated in a sequential fashion: $z^{s}_{1} \sim G_n$, and then for $j=2,\ldots,N$,
	\begin{equation*}
		\label{dp}
		z^{s}_{j}\mid z^{s}_{1},\ldots z^{s}_{j-1} \sim \frac{\alpha+n}{\alpha+n+j-1} G_n(\cdotp) + \frac{1}{\alpha+n+j-1}\sum_{k=1}^{j-1}\delta_{z^{s}_{k}}\left(\cdotp\right)\equiv G_{n+j-1}.
	\end{equation*}
	Each of the $S$ datasets generates a sampled variate from the posterior distribution by solving \eqref{MCminimizer} to yield $(\theta^{1},\ldots,\theta^{S})$, which, in the limit as $N \longrightarrow \infty$, is an exact sample from the posterior distribution for $\theta$. {Under this formulation, the difference between the prior-to-posterior approach via \eqref{wlikeInf} and the predictive-to-posterior approach via \eqref{MCminimizer} is that the latter integrates out the posterior DP and uses the collapsed form that relies only on sampling observables via the P\'{o}lya urn.}  Algorithm \ref{A1} implements the predictive-to-posterior via the P\'{o}lya urn scheme. Therefore, under exchangeability assumption, Algorithms \ref{A0} and \ref{A1} essentially yield the same posterior distribution via the Bayesian non-parameter modelling.

	\subsection{Verifying the validity of the DP-based loss calculation}
	\label{MBV}
	We investigate the validity of the proposed approach in Section \ref{Sec:postloss} using the Monahan and Boos approach discussed in Section \ref{sec:valid}.  Algorithm 1 in the Supplmentary Material details the computational strategy to verify the validity of our posterior. We first generate $\xi^{m}$ ($m=1,\ldots,M$) from a prior distribution $\pi_0(\xi)$, and for each $\xi^m$ we generate data $z_{1:n}^m$ from a parametric model $f(z|\xi^m)$.  Based on these data, we compute $H^{m}$ via \eqref{Hformula} with  $\theta=\theta^{m}$, the posterior sample obtained via \eqref{MCminimizer} based on $z_{1:n}^{m}$.  The collection of $\{H^{1},\ldots,H^{M}\}$ are used to assess uniformity.

	We illustrate the method using data simulated from the set up of Example 1 in Section \ref{Sec:sim}, which illustrates an application of a method of causal inference known as PS regression.  In this example, we generate the values of coefficients in the outcome model from the prior distribution, that is, from the Normal distribution with mean the same as the example and variance $10,000$, with the outcome data generated from a specific regression model. The loss function used to compute the posterior for the parameter of interest is specified as squared loss, based on a misspecified mean model that still allows consistent estimation of this parameter.  The procedure is repeated $1,000$ times with $n=100$ and $n=10,000$. For each dataset, we perform the proposed method for various $\alpha$ values, and produce the posterior sample of the ATE based on a correctly specified propensity score model and a misspecified outcome model that uses the treatment variable and the estimated propensity score only as covariates.
	
	Table \ref{kstest} displays the $p$-value of the Kolmogorov–Smirnov test for uniformity of the simulated $H$ for the causal parameter.  When $n=100$, the $p$-values suggest a proper posterior inference for all $\alpha=0,1,10$,  but when $\alpha=100$ this procedure fails the uniformity test. As when $\alpha$ becomes larger, there will be an increasing impact of model misspecification since the base measure for predictive inference is centered at the fitted value of the misspecified outcome model.  It will inflate the posterior variance to account for misspecification, and  this is confirmed  in the density plots in Figure \ref{MB}. The left panel of Figure \ref{MB} shows the density plots of $H$ with $n=100$, and we observe a higher dense around 0 and 1 when $\alpha=100$, demonstrating that the posterior variance is greater than those when $\alpha$ is smaller. However, as $n=10000$, the impact of model misspecification from the base measure becomes negligible, and  the $p$-values suggest coverage-valid posterior inference for all values of $\alpha$, which is confirmed by the density plots of the right panel of Figure \ref{MB}.  Note that, here, the inference model is misspecified compared to the data generating model, and yet the Monahan and Boos approach allows us to verify that the posterior credible intervals are valid in coverage terms.  Specifically, the fitted PS regression model -- which is misspecified by necessity -- still returns valid posterior coverage, even though its associated posterior distribution does not match the posterior distribution that would be obtained under correct specification of an outcome regression model (the posterior under correct specification would have smaller variance).

	We also produce the posterior distribution of the average treatment effect based on a misspecified PS model and a misspecified outcome model (with the treatment variable and the estimated propensity score).  In this case, the computed posterior will not concentrate at the true value as $n$ grows. In all cases, the posterior distribution computed using generalized Bayesian bootstrap inference fails the Monahan \& Boos uniformity test, with $p$-values all smaller than $10^{-6}$. This is confirmed by the density plots in Figure 1 in the Supplementary Material, where they exhibit higher densities at the tails,  demonstrating the impact of model misspecification and yielding a non-valid uncertainty inference.

	\begin{table}
		\caption{\small \label{kstest} $P$-values of  Kolmogorov–Smirnov test for uniformity of $H$ via Bayesian predictive inference with various  $\alpha$ values.}
		\renewcommand{\arraystretch}{0.5}
		\small
		\centering
		{
			\begin{tabular*}{30pc}{@{\hskip5pt}@{\extracolsep{\fill}}c@{}c@{}c@{}c@{}c@{\hskip5pt}}
				\hline
				$\alpha$ &	 0 &  1 & 10 & 100 \\
				\hline
				$n=100$ &  0.8632&0.4131& 0.0587 &  0.0000	\\
				$n=10000$ & 0.3998 & 0.6121& 0.4595&0.2917\\
				\hline
			\end{tabular*}
		}
	\end{table}

	\begin{figure}[ht]
		\centering
		\begin{minipage}[b]{0.45\textwidth}
			\includegraphics[width=0.8\textwidth]{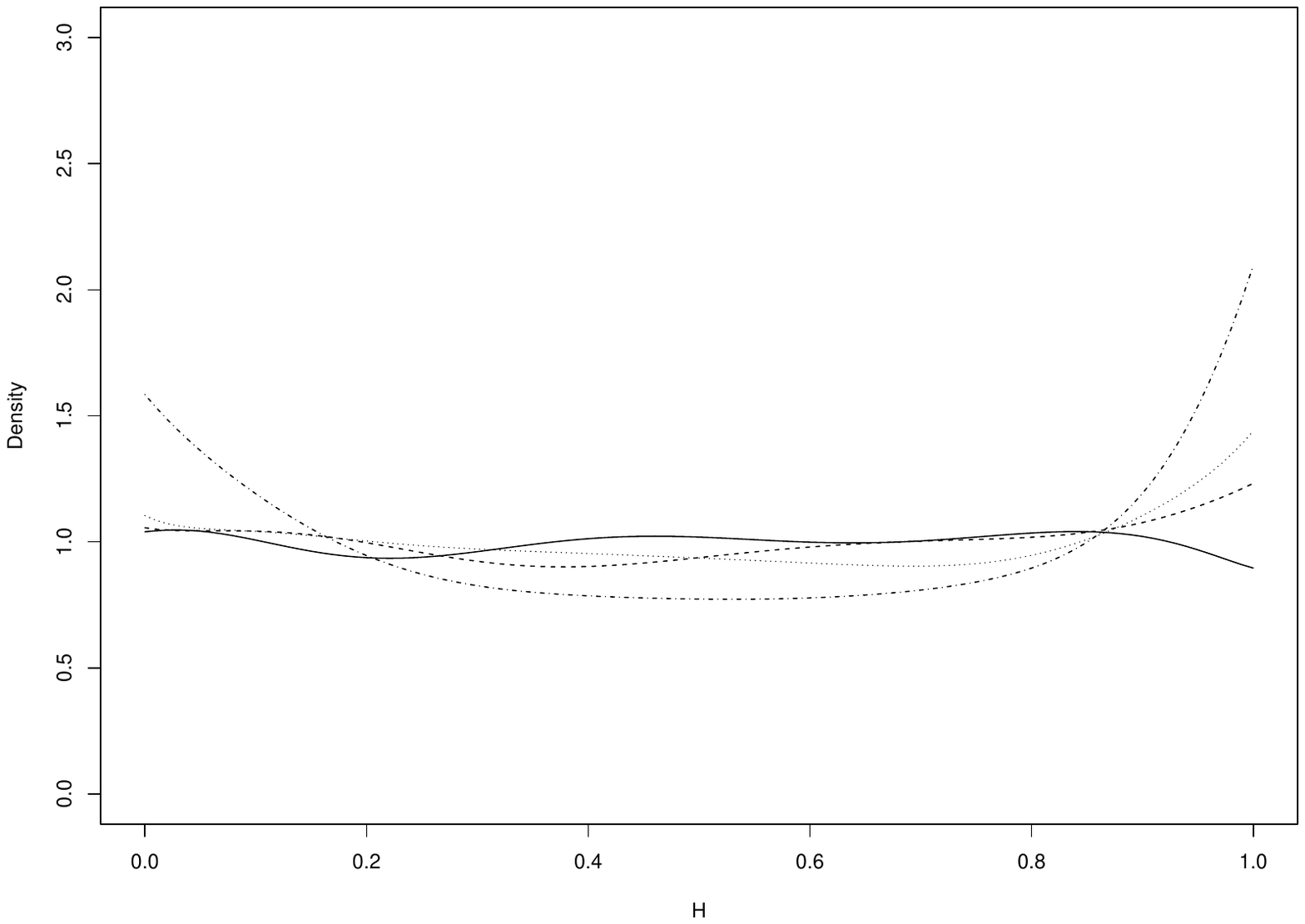}
		\end{minipage}
		\hfill
		\begin{minipage}[b]{0.45\textwidth}
			\includegraphics[width=0.8\textwidth]{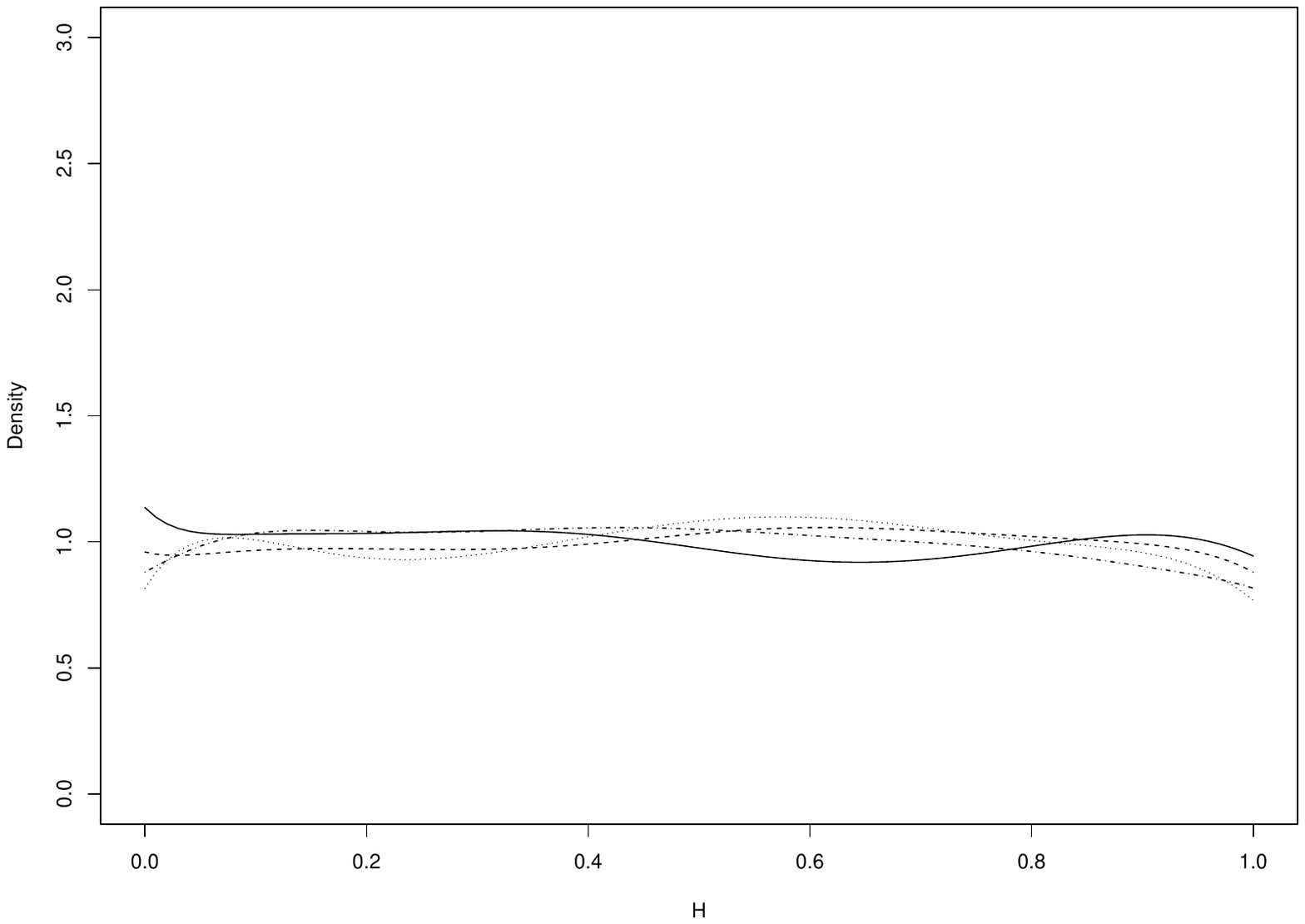}
		\end{minipage}
		\caption{\label{MB}    \small Density plots of $H$ with $n=100$ (left) and $n=10000$ (right). The solid, dashed, dotted  and dotted dash lines represent results from $\alpha=0,1,10,100$ respectively.}
	\end{figure}
	
	\subsection{Calibrating uncertainty via a valid posterior inference}
	Another inference mechanism can be derived directly using a loss function connecting the distribution of $Z$ and $\theta$ based on the definition in \eqref{u2samp}.  The approach derives a probability measure of $\theta$ to define the posterior distribution $\pi \left(\theta \left|z_{1:n}\right.\right)$ given a prior $\pi_0(\theta)$, and this \textit{Gibbs posterior} is defined as $\pi  (\theta | z_{1:n} )\propto \exp \left(-\eta \ell \left(z_{1:n},\theta\right) \right) \times \pi_0\left(\theta\right)$.  This posterior is defined if and only if the denominator is finite, and it gives a formal Bayesian procedure to update prior beliefs on $\theta$ to posterior beliefs based on the loss function and decision-theoretic arguments. In this Gibbs posterior approach, the scaling parameter $\eta$ has to be specified before performing inference. There are several proposals based on different criteria, such as the posterior coverage rate \citep{syring2019calibrating}.  Data-driven approaches for estimating $\eta$ have been studied extensively, as the issue of misspecification cannot guarantee that the posterior variance matches the sandwich variance obtained as in frequentist inference.   As demonstrated in \cite{chernozhukov2003mcmc}, under certain regularity conditions, the Gibbs posterior will be asymptotically normal with covariance matrix $\eta^{-1}(n\mathcal{J})^{-1}$ where $\mathcal{J}=- \E[\dot{\bU}(\theta^\ast)]$, $\mathbf{U}\left(\theta\right)={\partial \ell\left(z,\theta\right)}\big/{\partial \theta}$ and $\dot{\bU}(\theta) = {\partial \bU(\theta)}\big/{\partial \theta^\top}$. Therefore, asymptotically, the Gibbs posterior concentrates on a $\sqrt{n}$-ball centered at  $\theta^\ast$ with covariance matrix $\eta^{-1}(n\mathcal{J})^{-1}$. This result highlights that the learning rate is the key distinction between the Gibbs posterior and a misspecified Bayesian posterior \citep{kleijn2012bernstein}. In the Gibbs posterior construction, where it is necessary to balance the influence of the prior and the data, the scale of the loss function becomes important in practical applications. In the correctly specified case, the balance between prior and data contributions occurs naturally, resulting in both methods producing similar posterior variance. However, in misspecified cases, this balance is disrupted, and the variance of the Gibbs posterior requires manual adjustment via $\eta$ as it does not have the sandwich form. Failing to adjust $\eta$ can lead to posterior credible intervals with severely poor frequentist coverage probabilities \citep{martin2022direct}.  In Section \ref{MBV}, we verified that the Bayesian non-parametric modelling method for loss functions generate a  valid posterior, i.e., yielding a desire coverage rate. Therefore, we can calibrate $\eta$ based on posterior variance matching by first obtaining the sample posterior  variance $V$ from the sample generated by Algorithm \ref{A0} or \ref{A1}, and then find the proportional rate $c$ so that the two matrices are close in the norm sense, and therefore $\eta \approx c$ to achieve the similar uncertainty quantification. In practice, we do not know $(n\mathcal{J})^{-1}$ but can assess this value empirically. Algorithm \ref{A2} describes this algorithm in detail and find $c$ using Riemannian metric on the space of covariance matrices. 
	
	\begin{algorithm}[h]
		\renewcommand{\arraystretch}{0.5}
		\small
		\caption{\label{A2} \small Algorithm to calibrate $\eta$ by the DP-based loss posterior calculation.}
		\SetAlgoLined
		\KwData{$z_{1:n}=(z_1,\ldots,z_n)$}
		{\begin{itemize}[itemsep=0pt]
				\item Algorithm  \ref{A0} or \ref{A1} to obtain the posterior sample $(\theta^{1},\ldots,\theta^{S})$.
				\item Calculate the empirical posterior variance (or variance-covariance matrix) $V$.
				\item Obtain the Gibbs posterior sample from $\pi(\theta|z_{1:n})$ and calculate the posterior variance (or variance-covariance matrix)  $\hat \Sigma$ based on an initial guess $\eta_{0}$.
				\item Update $\eta$ according to $\hat \eta = \eta_0\times d$, where $d = ||\log V - \log \hat \Sigma||_{F}$, $\log$ is the matrix logarithm and $||\cdot||_{F}$ is the Frobenius norm.
			\end{itemize}
		}	
		\Return $\hat \eta $. 
	\end{algorithm}
	
	To verify the performance, we implemented an example from \cite{syring2019calibrating} that study problems concerning quantile regression.  In this example, $\theta=(\theta_0,\theta_1)=(2,1)$ is the coefficient, and the data are generated from $Y\sim \mathcal{N}(\theta_0 +\theta_1 X,1)$ and  $X+2 \sim \chi^2(4)$. The loss function is specified as the misspecified asymmetric Laplace likelihood, i.e., $\ell_n((y_i,x_i)_{i=1}^n,\theta)  = \frac{1}{n}\sum_{i=1}^{n} \left|(y_i-x_i^\top\theta)(0.5 - \mathbbm{1}_{(-\infty, x_i^\top\theta)}(y_i))\right|$. Table \ref{comp} shows the results for the marginal, joint coverage and the average length. Algorithm~\ref{A2} produces slightly wider confidence intervals and correspondingly higher marginal coverage probabilities than Algorithm~\ref{A1}. For both algorithms, the joint coverage is close to the nominal level. The method of \cite{syring2019calibrating} attains about nominal marginal coverage but undercovers in terms of joint coverage. These results indicate that the proposed learning-rate adjustment method achieves good performance despite its low computational cost.

	\begin{table}[ht]
		\caption{\label{comp} \small Comparison of 95\% posterior credible intervals from Algorithm \ref{A1} with $\alpha=0$, and Gibbs posterior calibrating using \cite{syring2019calibrating} and Algorithm \ref{A2}, based on 500 simulated datasets with $n=200$. SM represents the method proposed in \cite{syring2019calibrating}. }
		\renewcommand{\arraystretch}{0.5}
		\small
		\centering
		{\begin{tabular}{l*{1}{l}*{6}{c}}
				\hline
				&& \multicolumn{3}{c}{Coverage probability $\times$ 100}&   \multicolumn{3}{c}{Average length  $\times$  100} \\ \cline{3-8}
				&	&  Algorithm \ref{A1}	 & SM  & Algorithm \ref{A2} & Algorithm \ref{A1}	 & SM & Algorithm \ref{A2} \\
				\hline
				&	\multicolumn{7}{l}{Quantile regression example in Section 4 from \cite{syring2019calibrating}} \\			
				\hline
				&	$\theta_0$ & 96.0&  92.8& 97.8 & 71.1& 62.5& 93.5\\
				& $\theta_1$ & 97.0 & 94.9 & 99.2& 36.7 & 32.1& 49.3 \\
				& Joint    & 93.0 & 87.9 & 97.1  & -  & - & - \\
				\hline
		\end{tabular}}
	\end{table}
	\vspace{-0.2in}
	\section{Asymptotic results}
	
	\label{sec:AP}
	In this section, we establish the large sample properties of the proposed loss-type posterior calculation, under possible model misspecification. The required assumptions (which relate to identifiability, and regularity of the loss function) are classical, and proofs are included in the Supplementary Material. First, to show the consistency, we need to consider the limiting case as $n \longrightarrow \infty$. This requires posterior sample generated in Algorithm \ref{A0} and \ref{A1} to be degenerate at $\theta^\ast$ if all the information is available. Let $\pi_0(\theta)$ be the prior distribution for $\theta\in\Theta$.

	\begin{theorem}	
		Suppose the prior $\pi_0(\theta)$ has full Hellinger support, and $\ell\left(z,\theta\right)$ is continuous $\forall\theta \in \Theta$ with $$\int \log\left[1+\left|\ell\left(z,\theta\right) \right|\right]dG_0\left(z\right) < \infty.$$  Let $\theta^s_1 = \theta(\varpi^s,\zeta^{s})$ be the unique solution to
		\[
		\min\limits_{\theta \in \Theta} \sum_{k=1}^{\infty}\varpi^s_k \ell\left(\zeta^{s}_k,\theta\right)
		\]
		for any given $\varpi^s$ and $\zeta^s$ generated from Algorithm \ref{A0}.   Let $\theta^s_2=\theta^s(z^s)$ be the unique solution to
		$\min\limits_{\theta \in \Theta} \sum_{i=1}^{N} \ell\left(z^{s}_i,\theta\right)$
		for integer $N$, and any given $z^s$ generated from Algorithm \ref{A1}. Then
		\[
		\sum_{k=1}^{\infty}\varpi^s_k \ell (\zeta^s_k, \theta^s_1 )  \longrightarrow \min\limits_{\theta \in \Theta}\int \ell\left(z,\theta\right) d\Ftrue\left(z\right),\;\;\;\;\;
		\sum_{i=1}^{N}\ell (z^s_i, \theta^s_2 )  \longrightarrow \min\limits_{\theta \in \Theta}\int \ell\left(z,\theta\right) d\Ftrue\left(z\right)
		\]			
		and $\theta_1^s  \longrightarrow\theta^\ast$, $\theta_2^s  \longrightarrow \theta^\ast$ almost surely as $n,N \longrightarrow \infty$.
	\end{theorem}
	
	We can also consider the limiting behavior of the estimator in terms of the probability law, specifically, that it exhibits posterior asymptotic normality. In the empirical measure, the estimating equation becomes $\sum_{i=1}^n \bU_i(\theta) = \mathbf{0}$ define in \eqref{est} and with some regularity conditions, we have that the frequentist solution $\hat \theta_n$ has the property that
	$\sqrt{n}(\hat \theta_n - \theta^\ast) \xrightarrow{d} Normal_p(\mathbf{0},\bV)$, where $\bV = \mathcal{J}(\theta^\ast)^{-1} \mathcal{I}(\theta^\ast) \mathcal{J}(\theta^\ast)^{-\top}$
	with $\mathcal{I}(\theta^\ast) = \E[ \bU(\theta^\ast) \bU(\theta^\ast)^\top]$ and $\mathcal{J}(\theta^\ast) = -\E[\dot{\bU}(\theta^\ast)$, both $(p \times p)$ matrices, and $\dot{\bU}(\theta^\ast) = {\partial \bU(\theta)}\big/{\partial \theta^\top} \big|_{\theta = \theta^\ast}$.  The Bayesian analogy is the Bernstein-von Mises theorem, which establishes the limiting behavior of the posterior distribution.  We state the result in terms of the standardized parameter $\vartheta_{n,N}^s = \sqrt{N}( \theta^s -\hat \theta_n)$, where $ \theta^s $ is a draw from Algorithm \ref{A1}, and $\hat \theta_n$ is the frequentist estimator to \eqref{est}.

	\begin{theorem}
			\label{thm2}
			Under Assumptions 1-6 in Supplementary Material, the probability that the posterior for $\vartheta_{n,N}^s$ assigns to an arbitrary set $A\subseteq\Theta$ converges to the mass given by a Normal measure. Specifically, if $\mathbf{Z} \sim Normal_p(\mathbf{0},\bV)$ is an arbitrary random variable independent from all other random variables, then 
			$$\pi (\vartheta_{n,N}^s \in A\left|z_{1:n}\right. ) \to P(\mathbf{Z}\in A),$$
			 as  $n,N\to \infty$.
		\end{theorem}

		\section{Doubly robust causal inference via PS regression}
		\label{bayecausal}
		
		We now introduce our motivating example, which is a Bayesian representation for the DR regression approach to causal estimation.  In a causal inference setting, for the $i$th unit of observation, $Y_i$ denotes a response, $d_i$ the treatment (or exposure) received, and $x_i$ a vector of pre-treatment covariates or confounder variables. Suppose the {data generating structural} model is
		\begin{equation}
			\label{true}
			Y_i= \psi^\ast d_i + h_0(x_i) + \epsilon_i,\;\; \forall i=1,2,\ldots,n
		\end{equation}
		where $\mathbb{E}\left[\epsilon_i\left|x_i,d_i\right.\right] = 0$ and $\text{Var}\left[\epsilon_i\left|x_i,d_i\right.\right] = \sigma^2<\infty$, with $\epsilon_1,\ldots, \epsilon_n$ independent, and where $h_0(x_i)$ is an unknown real-valued function of the vector $x_i$. In this setting, $\psi^\ast$ is the ATE.

		\subsection{Frequentist inference in PS regression}
		
		{A typical approach to causal adjustment uses the PS.  With the PS estimated either via maximum likelihood or a fully Bayesian procedure summarized by the posterior mean, the outcome is modelled by adding the estimated PS \citep{robins1992estimating}, denoted $e\left(x_i;\hat\gamma\right)=\mathbb{P}\left(D_i=1\left|x_i;\hat \gamma\right.\right)$, where  $\gamma$ is estimated via some form of binary regression, or via more flexible prediction approaches. Assume we specify the augmented model as
			\begin{equation}
				\label{or}
				Y_i= \psi d_i + h_1(x_i) +  \phi e\left(x_i;\hat\gamma\right)+\epsilon_i,\;\; \forall i=1,2,\ldots,n
			\end{equation}
			and fit the model using ordinary least squares.  The model in \eqref{or} leads to doubly robust inference.  If  $h_1(x)=h_0(x)$, so that \eqref{or} matches \eqref{true} and the model is correctly specified, then the estimator of the true ATE will be consistent irrespective of whether the PS model is correctly specified because the estimator $\widehat \phi$ will converge to zero as $n \longrightarrow \infty$; on the other hand, if the PS is correctly modelled, conditioning on it will block the confounding path from $D$ to $Y$ via $X$ so that $X \perp \!\!\! \perp  D \left|\;e(X)\right.$, and \eqref{or} will still yield a consistent estimator of $\psi^\ast$, even if $h_1(x)$ is incorrectly specified.  We will proceed by assuming that the functional forms of $h_0(x_i,\beta_0)$ and $h_1(x_i, \beta)$ are parametric with associated parameter vectors $\beta_0$ and $\beta$. For example, linear regression assumes that $h_0(x_i,\beta_0)=x_i^\top\beta_0$ and $h_1(x_i,\beta)=x_i^\top\beta$.}

		Let $z_i = (y_i, d_i, x_i), i = 1, \ldots,n$, be the observed data, and $Z_i = (Y_i, d_i, x_i), i = 1, \ldots,n$ be the random variable representing the random component in the conditional model.  Estimation of $\theta=\left(\psi,\beta,\phi\right)$ in the conditional mean model \eqref{or} can proceed by defining a loss function which is the sum of squares of the residual error, i.e.,
		\begin{equation}
			\label{u1}
			\ell\left(z_{1:n},\theta\right)=\sum_{i=1}^{n} \left[y_i- \left(\psi d_i + h_1(x_i,\beta) +  \phi e\left(x_i;\hat\gamma\right)\right)\right]^2.
		\end{equation}
		The method does not make any distributional assumption about $\epsilon_i$,  and yields the solution
		\[
		\hat  \psi  = \frac{\sum\limits_{i=1}^{n}\left(d_i - e\left(x_i;\hat \gamma\right)\right) (y_i- h_1(x_i,\hat\beta)-\phi e\left(x_i;\hat \gamma\right))  }{\sum\limits_{i=1}^{n}\left(d_i - e\left(x_i;\hat \gamma\right)\right) d_i}.
		\]
		This is the feasible G-estimator \citep{robins1992estimating}, which is consistent for $\psi^\ast$ and robust to misspecification.  A key aspect of this frequentist approach is the use of plug-in estimation for parameter $\gamma$; it can be demonstrated that this approach provides locally efficient estimation of $\psi$ under the assumption that the PS model is correctly specified at least up to a finite dimensional parameter that may itself be estimated consistently at the usual parametric rate.  Non-parametric estimation of the PS model can also preserve consistent and efficient estimation of $\psi$, provided the rate of convergence of the non-parametric estimator is fast enough, and this can be achieved by using many standard flexible or machine learning (ML) approaches.

		\subsection{Bayesian inference in PS regression}
		
		The plug-in approach can also be justified in a fully Bayesian framework under the loss-based formulation. \cite{mccandless2010cutting} demonstrate how to block the flow of the information from the PS to the outcome regression when implementing MCMC in a joint model of the treatment and outcome. However, this approach induces finite sample bias and is not ideal for small sample inference.  A two-step approach, which assumes a complete separation in inference between the PS and outcome models and uses a plug-in estimate of $\gamma$ in \eqref{or} also yields a valid Bayesian solution \citep{stephens2021bayesian}.  This approach has been shown to provide superior estimation, and we adopt it in the following analysis.  The loss function used in \eqref{wlike} or \eqref{MCminimizer} should incorporate components for parameters in both the outcome model and the PS model, say $\ell(z,(\theta,\gamma)) = \ell_1(z,(\theta,\hat \gamma)) + \ell_2(z, \gamma)$ where $\hat \gamma$ is the minimizer of $\ell_2(\ldotp,\gamma)$ alone, with optimization over both sets of parameters carried out for each sampled realization from the proposed posterior distribution.
		
		We deploy the DP formulation from Section \ref{predpost}.  In the outcome setting, we assume that the predictive resampling is implemented through residuals arising from the outcome model \citep{quintana2020dependent}. In this case, we  first draw each  pair of $\left\{({x}^{s}_i,d^{s}_i)\right\}$, $i=1\ldots,N$, from the empirical distribution as the DP with $\alpha=0$, and then obtain the fitted values $e\left({x}^{s}_i;\hat \gamma^{s}\right)$ from a PS model based on logistic regression, refitted to the newly sampled $\left\{({x}^{s}_i,d^{s}_i)\right\}$ dataset. Then we simulate $y_i^s$ from a DP model with the conditional base measure $G_0 \equiv \mathcal{N}\left(\psi d^{s}_i+ h_1\left({x}^{s}_i,\beta\right)+ \phi e\left({x}^{s}_i,\hat \gamma^{s} \right),1\right)$, where  $\theta= \left(\psi,\beta, \phi\right)$ is generated from its prior distribution. 
		\begin{corollary}
			\label{thm3}
			The posterior distribution for the causal parameter, $\psi$ in \eqref{or}, becomes degenerate at $\psi^\ast$ as $n \longrightarrow \infty$ if either the outcome model or PS model is correctly specified. In addition, \textit{a posteriori}, $\theta  \xrightarrow{d} \mathcal{N}\left(\theta^\ast,V_{\theta^\ast}\right)$ where $V_{\theta^\ast} = \mathcal{J} (\theta^\ast)^{-1}\mathcal{I}(\theta^\ast)\mathcal{J} (\theta^\ast)^{-\top}$.
		\end{corollary}

		\section{Simulation studies}
		\label{Sec:sim}
		We examine the performance of the Bayesian methods described in Section \ref{Sec:postloss} with the two updating frameworks. For each example, we consider
		\begin{itemize}[nosep]
			\item Method I: Gibbs posterior, calibrating $\eta$ via posterior matching in Algorithm \ref{A2};
			\item Method II: Prior-to-posterior inference via the Bayesian bootstrap from \eqref{wlike};
			\item Method III: Generalized Bayesian bootstrap  method via Algorithm \ref{A0} or \ref{A1}.
		\end{itemize}

		\subsection{Example 1}
		In this example, we consider the simulation study constructed by \cite{saarela2016bayesian}. The data are simulated as {follows}:  we simulate $X_1,X_2,X_3,X_4 \sim \mathcal{N}\left(0,1\right)$ independently, and then set
		\begin{equation*}
			\begin{aligned}
				U_1 =\frac{\left|X_1\right|}{\sqrt{1-2/\pi}},\;\; D\left|U_1,X_2,X_3 \right. &\sim \text{ Bernoulli}\left(\text{expit}\left(0.4U_1+0.4X_2+0.8X_3\right)\right)\\
				Y \left|D,U_1,X_2,X_4\right. &\sim  \mathcal{N}\left(D-U_1-X_2-X_4,1\right)
			\end{aligned}
		\end{equation*}
		Three scenarios are considered:
		\begin{itemize}[nosep]
			\item Scenario A: Misspecify the OR  model using covariates $(x_1, x_2, x_4)$ and correctly specify a treatment assignment model using covariates $(u_1, x_2, x_3)$.
			\item Scenario B: Correctly specify the OR model using  covariates $(u_1, x_2, x_4)$ and misspecify a treatment assignment model using covariates $(x_1, x_2, x_3)$.
			\item Scenario C: Misspecify the OR model using covariates $(x_1, x_2, x_4)$ and misspecify a treatment assignment model using covariates $(x_1, x_2, x_3)$. This is not originally considered in \cite{saarela2016bayesian}.
		\end{itemize}
		
		The prior-to-posterior update via the Gibbs posterior is implemented using MCMC and the Bayesian bootstrap approaches, and non-informative priors are placed for all the parameters with $10,000$ MCMC samples and $1,000$ burn-in iterations.  For the predictive-to-posterior update, we generate $S=1,000$ sets, each with $N=10,000$ new data points and with $\alpha=1$. For $n=20$,  we also place an informative normal prior with mean {at the true value} and standard deviation 2 for the Gibbs posterior using MCMC.
		
		The results are given in Table~\ref{sim1}, and the table shows the results of $1,000$ Monte Carlo replicates of the averages of the posterior means, variances and coverage rates for $\theta$ with different sample sizes.  Coverage rates are computed by constructing a 95\% credible interval for $\theta$ from the 2.5\% and 97.5\% posterior sample quantiles.   When the sample size is small, the Bayesian bootstrap (Method II) and predictive inference models (Method III) exhibit poor coverage while the Gibbs posterior (Method I) returns coverage rates at the nominal level; however, Method II presents rather larger variances. The results for the Gibbs posteriors with $\eta=1$ display the correct posterior mean, but the coverage is significantly below the nominal level in Scenarios A and B, which confirms that the calibration of $\eta$ is required. The difference in variances diminishes as the sample size increases, or when the informative prior is considered (demonstrated in the bracket for $n=20$). The Bayesian bootstrap and DP-based predictive inference generate similar results when the sample size is over 100, as the prior does not carry much weight when $\alpha/N$ is small.  When both models are misspecified, all cases yield significantly biased estimates unless the informative prior is used.
		
		\begin{table}[ht]
			\renewcommand{\arraystretch}{0.5}
			\caption{\label{sim1} {Example 1: Simulation results of the marginal causal contrast, with true value equal to 1,  for $1,000$ simulation runs on generated datasets of size $n$. Gibbs represents results generated from the Gibbs posterior with $\eta=1$. The bracketed results are from the informative normal prior.}}
			\centering
			\small
			\begin{tabular*}{40pc}{@{\hskip5pt}@{\extracolsep{\fill}}l@{}c@{}c@{}c@{}c@{}|c@{}c@{}c@{}c@{}|c@{}c@{}c@{}c@{\hskip5pt}}	
				\hline
				& \multicolumn{4}{c}{Scenario A}&\multicolumn{4}{c}{Scenario B}&\multicolumn{4}{c}{Scenario C} \\ \cline{2-13}
				$n$	&20 &50 & 100 &500	&20  &50 & 100 &500	& 20  &50 & 100 &500	\\
				\hline	
				\multicolumn{13}{l}{Mean}	\\
				\hline	
				\multirow{2}{*}{Method I} &0.98  & 0.98& 1.01& 1.00 &0.98  &1.01 &1.00& 1.00&0.62 &0.77& 0.62 & 0.61\\
				&(1.05)& --&-- &--& (1.03)& --&-- &-- & (0.79)& --&-- &--\\
				Method II & 0.93 &0.99&1.00 &1.00  & 1.13&1.01&1.00& 1.00& 0.45& 0.63&0.64&  0.63\\
				Method III  & 0.95 &1.01&0.99&1.00&1.01 &0.99&0.99&1.00& 0.60&0.63&0.62& 0.62\\
				Gibbs   & 1.08& 0.99& 1.00& 1.00  &0.88& 0.99 &1.00& 1.00 &0.86& 0.62& 0.64 & 0.63\\
				\hline
				\multicolumn{13}{l}{Variance} \\
				\hline	
				\multirow{2}{*}{Method I} &1.12& 0.13&0.06&0.02 &1.23 &0.12&0.05& 0.01&0.52& 0.20&0.09&0.02 \\
				&(0.35)& --&-- &-- & (0.32)& --&-- &--& (0.22)& --&-- &--\\
				Method II &  13.52  &0.11&0.06 &0.01  &8.40 &0.12&0.05& 0.01& 69.89& 0.23 &0.09&0.02\\
				Method III& 0.46& 0.13&  0.05&0.01&0.39&0.12 &0.05&0.01& 0.68&0.19&0.09 &0.02\\
				Gibbs & 0.49& 0.13&  0.06&0.01&0.37& 0.12 &0.05 &0.01& 0.68&0.13&0.11 &0.02\\
				\hline
				\multicolumn{13}{l}{Coverage, \%}	\\
				\hline	
				\multirow{2}{*}{Method I} &96.5 & 95.5 & 95.1& 94.9 &94.9  &95.0 &95.0& 95.4& 93.2 &91.2& 85.3 & 25.6\\
				&(97.3)& --&-- &--& (95.9)& --&-- &-- & (95.6)& --&-- &-- \\
				%Method II &94.3 & 95.2 & 94.6& 95.3 &95.7  &95.0 &95.0& 95.4& 91.7 &88.4& 80.6 & 22.3\\
				Method II& 89.3 &92.2 & 93.5&  94.2& 82.2& 89.9&92.7&94.8 & 77.4 & 77.2& 74.2& 20.9\\
				Method III& 92.2 & 95.0 &  94.9 & 93.8  &  85.4  & 91.5 & 94.0 &  94.7  &77.5 &81.4&76.8&35.4\\		
				Gibbs  &	84.4&79.1 & 80.1& 84.6 &78.3&84.8 & 84.0 & 84.3 & 66.7&52.5& 42.7 & 4.2			 \\
				\hline
			\end{tabular*}
		\end{table}

		\subsection{Example 2: High-dimensional case}

		In this example, we examine the performance of the proposed updating approaches under high  dimensional settings, with binary exposure.  The data are simulated as follows: we simulate $X=(X_1,X_2, \ldots,X_{p}) \sim \mathcal{N}_p\left(0,\Sigma\right)$ and $\Sigma_{ij} =1$ if $i=j$ and 0.1 otherwise, and then simulate
		\begin{equation*}
			\begin{aligned}
				D\left|X \right. &\sim \text{ Bernoulli}\left(\text{expit}\left(0.45X_1+0.9X_2-0.4X_5+1.3X_2X_5+1.8X_1X_2\right)\right)\\
				Y \left|D,X\right. &\sim  \mathcal{N}\left(D+0.5X_1+X_3-0.1X_4-0.2X_7+ 1.5X_3X_4 +0.6X_7^2 +1.2X_1X_3,1\right).\\
			\end{aligned}
		\end{equation*}
		In the analyses, we take $p=20$ and $n=50$ and $100$. The loss function adopted for the loss-based analysis for Method II and Method III encompasses both the need to penalize the number of terms in the PS model and the need to select a penalization parameter.  We use penalized logistic regression for the PS model including all $x_1,\ldots,x_p$ and first order interactions between them, giving a total of $q=p+p(p-1)/2$ parameters.  Specifically, we use the lasso penalty, and base conclusions on the loss function $ \ell_2((x,d), (\gamma,\lambda)) = -\log f_{D|X}(d|x,\gamma) +\lambda \sum_{j=1}^{q}\left|\gamma_j\right|
		$, where $f_{D|X}(d|x,\gamma)$ is the Bernoulli mass function with logistic link.  This loss function is then incorporated into a cross-validation procedure to define the loss to be deployed in the implementation of Methods II and III, $\ell_{\text{CV}}((x,d),(\gamma,\lambda))$ say, which takes the input data and returns optimized values of $\gamma$ and $\lambda$, as well as the fitted values that can be transported into the outcome model. The value of $\lambda$ is estimated with lowest test mean squared error over 10-fold cross validation. For the {outcome model}, we fit the model with the treatment indicator and estimated PSs only as covariates. For Method II, we set $S=1000$. For Method III, we generate $S=1000$ datasets, each with $N=100$ and $\alpha=5$.  For comparison, we also include the Bayesian doubly robust high-dimensional (BDR-HD) method of \cite{antonelli2022causal}, in which the PS and outcome models are estimated using Gaussian process (GP) priors and the resulting MCMC estimates are plugged into a DR estimator. The variance is adjusted through the frequentist bootstrap so that it will achieve frequentist nominal coverage rate. For the BDR-HD method, we ran 500 iterations with 100 burn-in iterations for both the PS and OR models.
		
		The results are given in Table \ref{sim2s}. Method II shows the smallest bias among all methods, while Method III both exhibit some biases in $n=50$ and $n=100$; this is primarily due to the bias from the fitted PS via the lasso penalty. Additionally, the outcome model specified in Methods II and III only contains the PS and an intercept to account for confounding. However, the coverage rate is still around the nominal level for both methods. If the PS is more accurately estimated, the bias would diminish and the coverage rate would achieve the target level as suggested by the additional simulation study in Section 3 of the Supplementary Material. Method III shows a larger bias due to the impact of  the new data which are generated from a misspecified model. The BDR-HD method exhibits small biases in both cases, and the coverage rates are around nominal level. In this method, it utilizes component-wise GP regression for each confounder in treatment and outcome models, and interaction terms are supplied in the GP regression as separate covariates. Therefore, it achieves desired performance. However, in practice, this might not be feasible as it will include $p(p-1)/ 2$ additional covariates, increasing the computational complexity by at least $O(p^2)$.  It should be noted that  BDR-HD is substantially more computationally intensive than Methods II and III, as it requires MCMC for GP regression and an additional bootstrap step to adjust posterior variance, resulting in much longer run times per replicate.

		\begin{table}
			\caption{\label{sim2s} \small Example 2: Simulation results of the ATE under high-dimensional settings, with true value equal to 1, based on 500 simulation runs. BDR-HD represents the method proposed in \cite{antonelli2022causal}. Running time is the average time per Monte Carlo replicate in minutes.}
			\centering
			\small
			\renewcommand{\arraystretch}{0.5}
			\begin{tabular}{c c c c c c}
				\hline
				$n$ & Method & Bias & RMSE & Coverage rate & Running time \\
				\hline
				\multirow{3}{*}{50}
				& Method II   & 0.092 & 0.818 & 98.8 & 1.87 \\
				& Method III  & 0.253 & 0.634 & 93.4 & 1.81 \\
				& BDR-HD      & 0.232 & 0.728 & 98.0 & 15.19 \\
				\hline
				\multirow{3}{*}{100}
				& Method II   & 0.037 & 0.576 & 98.8 & 2.51 \\
				& Method III  & 0.306 & 0.389 & 97.0 & 1.79 \\
				& BDR-HD      & 0.112 & 0.343 & 97.8 & 32.36 \\
				\hline
			\end{tabular}
		\end{table}
		
		\subsection{Example 3: Comparison with flexible modelling approaches}
		In this example, we seek to compare the proposed approach with existing flexible causal estimation approaches.  We compare Method III with Bayesian causal forests (BCFs,  \cite{hahn2020bayesian}), and double ML (DML) \citep{chernozhukov2018double} using a variety of ML strategies.  We first consider the following data generating mechanism with interaction terms:	
		\begin{equation*}
			\small
			\begin{aligned}
				X_1,X_3&\sim \mathcal{N}\left(1,1\right), X_2,X_4 \sim \mathcal{N}\left(-0.5,1\right) \\
				D\left|X_1,X_2,X_3,X_4 \right. &\sim \text{ Bernoulli}\left(\text{expit}\left(0.3X_1+0.9X_2-1.25X_3+1.5X_4\right)\right)\\
				Y|D,X_1,X_2,X_3,X_4 &\sim  \mathcal{N}(\mu_0(D,X; \beta), 1 )\\
				\mu_0(D,X) =   10+D (1+ 2X_1) &+ X_1+ X_2 + X_3  +X_4 + 0.25X_1^2 + 0.75 X_2 X_4 + 0.75X_3 X_4 . 
			\end{aligned}
		\end{equation*}
		
		The ATE then is $\E[ \mu_0(1,X) ] - \E[ \mu_0(0,X) ] = 1 +  2\E[X_1] = 3$. Since there are interaction terms in the OR, we specify the  mean of the treatment-effect model as $\beta + (\theta+\theta_1x_1)d + (\phi_1 +\phi_2 x_1 )e\left(x;\hat\gamma\right)$. This model yields a consistent estimate for the ATE, and is fitted via the square loss through Method III with $\alpha=5$. We also consider the BCF method in \cite{hahn2020bayesian}. The BCF is a flexible approach for the outcome mean model using the Bayesian additive regression trees (BARTs) to infer the individual treatment effects, and it is based on linear predictor $\mu(d,x) = h(x,e\left(x;\hat\gamma\right)) + t\left(x,e\left(x;\hat\gamma\right) \right)d$, with assumed normal errors. The functions $ h(\cdot,\cdot) $ and $ t(\cdot,\cdot) $ are estimated via the BCFs. In this analysis, we assume the PS model is correctly specified and estimated via a parametric logistic regression in BCFs and the proposed approach.  Finally, we consider a frequentist DML approach. In this method, the ATE estimator, $\theta$, is the solution to $ \E [\psi(Z;\theta, \mu,e(X))]=0$, where   $\psi(\cdot)$ is the Neyman-orthogonal moment equation and defined as
		\[
		\psi(Z;\theta,\mu,e(X))= \mu(1,X)-\mu(0,X) +\frac{D(Y-\mu(1,X))}{e(X)} - \frac{(1-D)(Y-\mu(0,X))}{1-e(X)} - \theta
		\]
		and $\mu(\cdot,\cdot)$ is the treatment-effect model and $e(\cdot)$ is the propensity score. Both of them are estimated via various ML approaches. Specifically, we use the DML estimator in Definition 3.2 in \cite{chernozhukov2018double}, which the data are partitioned into $K$ groups.   The functions $\hat \mu_k(\cdot,\cdot)$ and $\hat e_k(\cdot)$ are estimated using the all the data excluding the $k$th group. Then the DML estimator for the ATE  is the solution to $ 1/K \sum_{k=1}^{K}\E_k [\psi(Z;\theta,\hat \mu_k,\hat e_k(X))]=0$, where $\E_k(\cdot)$ is the empirical expectation over the $k$th fold of the data.
		
		Table \ref{flex} displays the results of this study. 	In the DML framework, we estimate the OR $\mu(\cdot,\cdot)$ and the PS $e(\cdot)$ using several ML methods, including regression trees (CART), random forests, tree-based boosting, and neural networks (with two neuros). We also consider two hybrid approaches. The `Ensemble' method combines boosting, random forests, and neural networks, while the `Best' method selects, for each of $\mu(\cdot,\cdot)$ and $e(\cdot)$, the estimator that minimizes the average out-of-sample prediction error for the ATE among the candidate methods. BCF exhibits relatively low variance and consequently smaller RMSEs than Methods II and III. In contrast, all DML estimators display substantial bias that does not diminish with increasing sample size; combined with their relatively large variances, this leads to higher RMSEs overall. As for the coverage rate, DML coverage rates deteriorate rapidly as $n$ increases and eventually fall below the nominal level, except the random forest, which maintains near-nominal coverage. BCF consistently undercovers, whereas Method III achieves nominal coverage across all scenarios. Although BCF and DML do not require a specified functional form for the treatment effect model, unlike PS regression, both approaches are substantially more computationally intensive than Method II and III.

		\begin{table}[h]
			\caption{\label{flex} \small Comparison of results for the proposed Bayesian empirical likelihood, Bayesian causal forests (BCFs) and frequentist double machine learning estimator. Summary of $1,000$ simulation runs. Rows correspond to the bias, root mean square error (RMSE), and coverage rates.}
			\centering
			\small
			\renewcommand{\arraystretch}{0.5}
			\begin{tabular*}{40pc}{@{\hskip5pt}@{\extracolsep{\fill}}c@{}c@{}c@{}c@{}c@{}c@{}c@{}c@{}c@{}c@{}c@{\hskip5pt}}
				%\begin{tabular}{c|ccccccccc}
				\hline
				$n$ & & Method II & Method III & BCF & Tree & Forest & Boosting & Nnet & Ensemble & Best \\
				\hline
				\multirow{3}{*}{200} 
				& Bias       &-0.128& -0.007 & 0.021 & -0.348 &  0.252 &-0.147&     -0.089& -0.022& -0.111  \\
				& RMSE       & 0.322& 0.339 & 0.293 & 0.448 &   0.401 & 0.344 &   0.341 & 0.317& 0.346  \\
				& Coverage   & 88.8 & 89.9 & 92.3  & 91.0  & 93.2 &93.6   &   95.1    &96.0 &  94.4    \\
				\hline
				\multirow{3}{*}{1000} 
				& Bias       & -0.000&-0.007 & -0.004 & -0.366&   0.060&-0.116 &    -0.089&  -0.080& -0.092
				\\
				& RMSE       & 0.142&  0.137 & 0.114 & 0.390& 0.153 & 0.179 &  0.168& 0.158 &  0.169   \\
				& Coverage   & 89.7 & 95.8  & 90.8  & 42.3  &  94.0& 90.8    &   92.0    &92.8 &  92.0 \\
				\hline
				\multirow{3}{*}{2000} 
				& Bias       & 0.001 & 0.001 & 0.001 & -0.255 &   0.013& -0.102&    -0.088&  -0.067&  -0.089 \\
				& RMSE       & 0.100 &0.101 & 0.080 & 0.272 &  0.100&0.139&  0.134& 0.116& 0.135   \\
				& Coverage   & 91.1 & 94.5  & 89.5  & 45.4   & 95.4&  85.2&     86.4  & 90.7  &  86.0    \\
				\hline
			\end{tabular*}
		\end{table}

		\subsection{Application: UK Speed Camera Data}
		\label{Sec:app}
		Our real example aims to quantify the causal effect of speed camera presence on road traffic collision. We use data on the location of fixed speed cameras for 771 camera sites in the eight English administrative districts, including Cheshire, Dorset, Greater Manchester, Lancashire, Leicester, Merseyside, Sussex and the West Midlands. These data form the `treated' group. For the `untreated' group, we randomly select a sample of 4,787 points on the network within our eight administrative districts. Details of these data can be found in \cite{graham2019speed}. 	The outcome of interest is the number of personal injury collisions per kilometer.  The data are taken from police reports collated and processed by the Department for Transport in the UK in the `STATS 19' dataset. The location of each personal injury collision is recorded using the British National Grid coordinate system and can be located on a map using Geographical Information System  software. Data are collected from 1999 to 2007 to ensure the availability of collision data for the years before and after the camera installation for every camera site as speed cameras were introduced varying from 2002 to 2004. There is a formal set of {location} selection guidelines for speed {cameras} in the UK \citep{gains2004national}.  These {guidelines inform the selection of} covariates which represent the characteristics of units that simultaneously determine the treatment assignment (camera location) and outcome {(number of accidents)}. Primary guidelines for site section include the site length, the number of fatal and serious collisions and the number of personal injury collisions {in a preceding time period}. In addition, drivers might try to avoid the routes with speed cameras, and the reduction in collisions may come from a reduced traffic flow. Therefore, we include the annual average daily flow as a confounder to control the effect due to the traffic flow.  We also include factors that would have additional safety impacts, such as road types, speed limit, and the number of minor junctions within site length \citep{christie2003mobile}.
		
		We apply the proposed Bayesian methods to the speed camera data with the loss defined in \eqref{u1}.  \cite{graham2019speed} estimated the PS with a generalized additive model by including smooth functions on the annual average daily flow and the number of minor junctions and achieved balance and overlap. For the outcome model, we include all the confounders and the estimated PS.  We place non-informative priors for all the parameters in prior. For Method III, we set $\alpha =100$ and $N=10,000$. Table~\ref{app} shows summary statistics of the ATE based on $20,000$ posterior samples.  All methods indicate that the installation of a speed camera can reduce road traffic collisions, by approximately 1.4 incidents per site on average; however, we notice that Method II yields a slightly higher variation. The Gibbs posterior (Method I) has similar variance to the other two methods when calibrated using Algorithm \ref{A2} with $\eta = 0.024$, which is close to the calibration achieved by the estimated residual variance ($0.027$).  Table~\ref{app} also includes the posterior predictive distribution of the percentage reduction in the average change in road traffic collisions attributable to speed cameras.  PS regression–based methods estimate an approximately 18\% reduction in road traffic collisions at locations with speed cameras, suggesting a stronger causal effect than that obtained from Bayesian inverse probability weighting (IPW). All three loss-bases methods show similar posterior densities for the change in the ATE (Figure 2 in the Supplementary Material), while Method III shows a slightly smaller variance.   Unlike IPW, which relies solely on weighting, PS regression includes a treatment-free component, providing extra robustness when at least one model is correctly specified. Regression-based methods combined with the our proposed Bayesian calculation reduce the impact of extreme PS, yielding narrower 95\% credible intervals for the ATE.
		
		\begin{table}
			\caption{\label{app} \small Summary statistics for the posterior distribution of the ATE,  and the posterior predictive distribution of the percentage change of the ATE for speed camera data. IPW-BB represents results using the two-step Bayesian bootstrap approach based on inverse probability weighting estimation, while  IPW-BB (plug-in) represents the plug-in approach using the two-step Bayesian bootstrap.}
			\centering
			\small
			\renewcommand{\arraystretch}{0.5}
			{\begin{tabular*}{35pc}{@{\hskip5pt}@{\extracolsep{\fill}}c@{}c@{}c@{}c@{\hskip5pt}}
					\hline		
					&Posterior Mean & Standard Deviation & 95\% Credible Interval	\\
					\hline		
					\multicolumn{4}{l}{\textit{ATE}}\\
					Method I& -1.413& 0.183 & (-1.771, -1.054)\\
					Method II & -1.411 & 0.184 & (-1.772, -1.048)\\
					Method III & -1.413 & 0.180 & (-1.767, -1.058)\\
					IPW-BB & -1.089 & 0.203 & (-1.486, -0.679)\\
					IPW-BB (plug-in) & -1.088 &  0.209 & (-1.484, -0.663)\\
					\hline
					\multicolumn{4}{l}{\textit{Percentage Change of the ATE}}\\
					Method I &-18.656& 2.356 & (-23.250, -13.996) \\
					Method II  & -18.603& 2.352 & (-23.224, -13.951)\\
					Method III & -18.659& 2.313 & (-23.194, -14.070)\\
					IPW-BB & -14.338& 2.622 & (-19.419, -9.036)\\
					IPW-BB (plug-in)& -14.625& 2.807 & (-19.978, -8.877)\\
					\hline
				\end{tabular*}
			}
		\end{table}

		\section{Discussion}
		\label{Sec:dis}
		
		We have developed a formal Bayesian framework for inference on parameters defined via loss functions, without relying on the standard prior–likelihood formulation. While traditional Bayesian updating provides coherent decision-making under uncertainty, predictive inference allows quantification of uncertainty using a sequence of predictive distributions rather than a prior–likelihood specification. This approach often improves computational efficiency by relying on optimization instead of MCMC integration. Our focus is on non-parametric methods based on the DP. First, using the traditional Bayesian updating approach, we obtained the posterior distribution from a non-parametric prior–likelihood perspective. Second, from a Bayesian decision-theoretic viewpoint, we computed the posterior by minimizing a loss function over sequentially imputed sets of exchangeable data. These porposed approaches can naturally extend to multi-parameter problems, such as two-stage causal inference methods, overcoming a limitation of the Gibbs posterior. We also demonstrated that computations following this paradigm produces valid posterior inference in the spirit of \cite{monahan1992proper}. Simulation studies show that the proposed approaches possess strong Bayesian and frequentist properties while being computationally more efficient than other existing Bayesian methods. We applied the loss-based framework to assess the causal effect of speed cameras on road traffic accidents, finding that their presence reduces car-related collisions. Such inference can help transportation authorities design more effective, targeted installation plans to improve road safety. More generally, Bayesian methods provide interpretable uncertainty estimates in finite-sample causal inference applications.

		The principles introduced in this paper can be extended to much broader settings where likelihood functions are unavailable or difficult to specify. In such cases, the loss-based Bayesian framework provides a flexible alternative for conducting inference without relying on a fully specified likelihood. Moreover, the methodology can be applied to a wide range of causal inference problems when the set-up requires over-specifying the model condition. Future research could explore extensions to high-dimensional and longitudinal causal models, dynamic treatment regimes, and network-dependent data, as well as the integration of non-parametric priors with ML-based predictive models to further enhance computational efficiency and robustness.

\section*{Code Availability}
The code for the simulation studies is publicly available at  \url{https://github.com/yumcgill/Bayes-Loss}.

\bibliographystyle{chicago}
\bibliography{Bib-BayesCausal.bib}

\FloatBarrier
\clearpage

\appendix

\begin{center}
	{\LARGE\bf Supplementary material for ``Non-parametric Bayesian inference via loss functions under model misspecification''}
\end{center}

\section{Sampling from the Dirichlet process}
The $\{\varpi_k\}$ can also be generated to be monotonically decreasing in $k$ using an algorithm designed to simulate a Gamma process \citep{walker2000miscellanea}.  If $\{U_k\}$ are a sequence of independent $Uniform(0,1)$ random variables, we define $T_1 = h^{-1}(-\log U_1)$ and $T_k = h^{-1}(h(T_{k-1})-\log U_k)$ for $k = 2,3,\ldots$, where
\[
h(t) = \alpha \int_t^\infty \frac{1}{x} e^{-x} \ dx
\]
is the exponential integral, a monotonic decreasing function of $t$, with $h(0) = \infty$ and $h(t) \rightarrow 0$ as $t \rightarrow \infty$.  Then $
\varpi_k = T_k/\sum_{j = 1}^\infty T_j$ for $k=1,2,\ldots$ form a series of monotonically decreasing probabilities.  The quantities $\{T_k\}$ are themselves monotonically decreasing, allowing straightforward evaluation of the infinite sum up to machine precision.  Key relevant references include \cite{muliere1996bayesian,muliere1998approximating,ishwaran2002exact}.

\section{Assessing coverage validity}
Algorithm \ref{AMB} details the computational strategy to verify the validity of our posterior using the \cite{monahan1992proper} approach.
\begin{algorithm}[h]
	\renewcommand{\arraystretch}{0.5}
	\caption{\label{AMB} Algorithm to implement \cite{monahan1992proper} for assessing coverage validity.}
	\SetAlgoLined
	\SetKwInOut{Input}{Input}
	\SetKwInOut{Output}{Output}
	\Input{Data generating model $\ftrue(z|\xi)$ and prior $\pi_0(\xi)$}
	\For{$m=1,\ldots,M$}{
		{\begin{itemize}[itemsep=0pt]
				\item Simulate $\xi^m \sim \pi_0$;
				\item Simulate data $z_{1:n}^m \sim \ftrue(z|\xi^m)$;
				\item Compute $\theta^{\ast m}$ from
				\begin{equation}
					\label{u2alg}
					\theta^{\ast m} = \arg\min\limits_{\theta\in \Theta}\int \ell\left(z,t\right) d\Ftrue\left(z|\xi^m \right);
				\end{equation}
				\item Compute the proposed posterior $\widetilde \pi(\theta|z_{1:n}^m)$;
				\item Produce a posterior sample of size $N$, $\theta_{1}^m,\ldots,\theta_{N}^m$ from $\widetilde \pi(\theta|z_{1:n}^m)$;
				\item Record
				\[
				H^m = \frac{1}{N} \sum_{i =1}^N \mathbb{1}(\theta_i^m \leq \theta^{\ast m});
				\]
		\end{itemize}}
	}	
	\Return Test of uniformity of $(H^1,\ldots,H^M)$. 
\end{algorithm}

Figure \ref{MB1} shows density plots of $H$ with $n=100$ (left) and $n=10000$ (right) based on a mis-specified propensity score model and a mis-specified outcome model (with the treatment variable and the estimated propensity score). 
\begin{figure}[ht]
	\begin{minipage}[b]{0.45\textwidth}
		\includegraphics[width=0.8\textwidth]{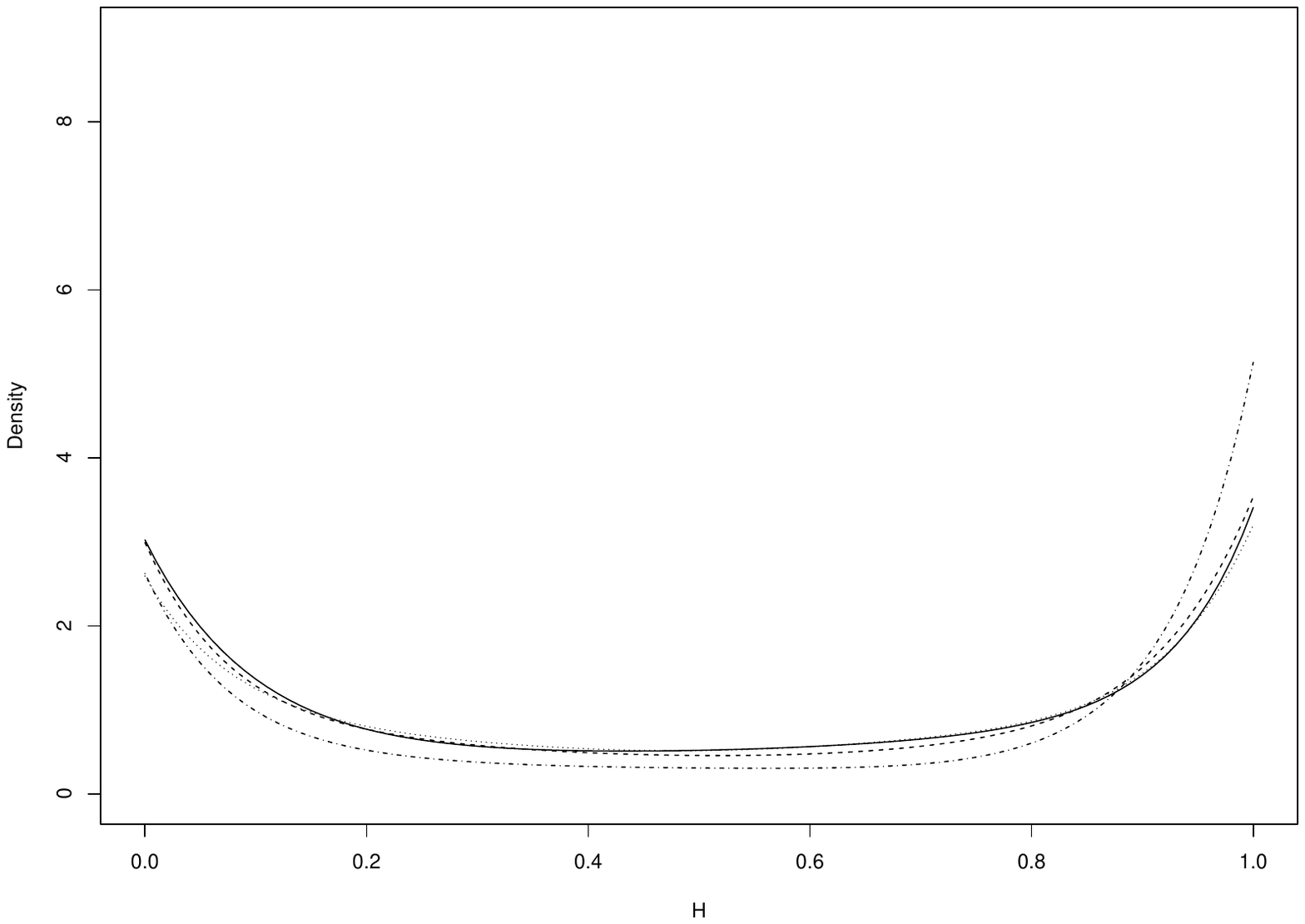}
	\end{minipage}
	\hfill
	\begin{minipage}[b]{0.45\textwidth}
		\includegraphics[width=0.8\textwidth]{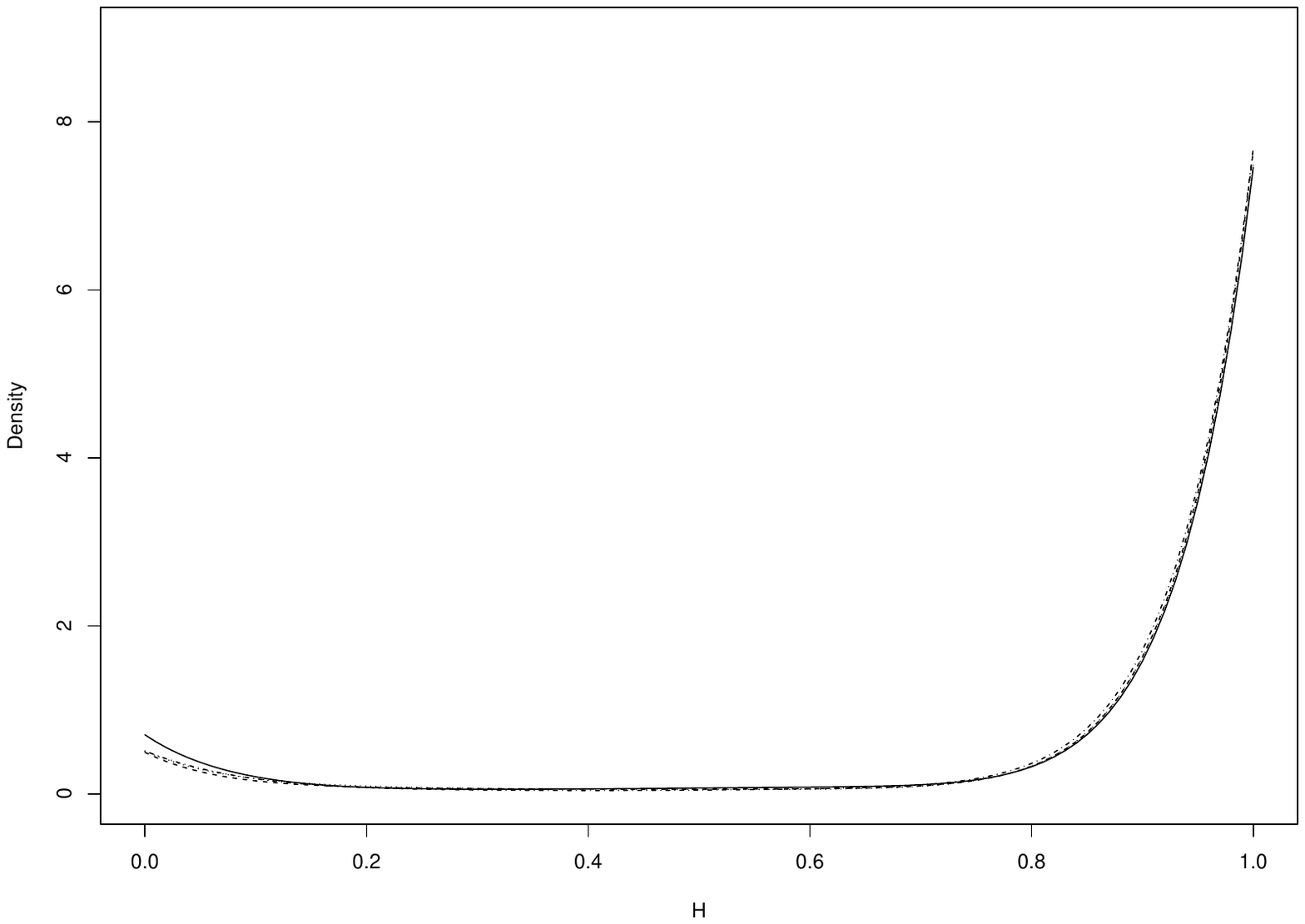}
	\end{minipage}
	\caption{\label{MB1} Checking coverage validity using Monahan \& Boos: Density plots of $H$ with $n=100$ (left) and $n=10000$ (right) under a mis-specified model. The solid, dashed, dotted  and dotted dash lines represent results from $\alpha=0,1,10,100$ respectively.}
\end{figure}

\section{Estimation via predictive inference and the KL divergence}
The Bayes estimator of the a target parameter is the function of the data that minimizes the posterior expected loss, given by
\begin{equation*}
	\arg \min\limits_{\theta \in \Theta} \int_{\Xi} u\left(\theta,\xi \right) \pi\left(\xi | z_{1:n} \right) \ d \xi.
\end{equation*}
If $u$ is taken to be the KL divergence between the true model, $\ftrue$ and a possibly mis-specified model, $f$, given by
\[
u\left(\theta, \xi\right)= \int  \log\left(\dfrac{\ftrue(z | \xi )}{f(z | \theta )}\right) \ftrue(z | \xi )  \ dz,
\]
then the optimization becomes
\begin{equation}
	\label{pp1}
	\arg \min\limits_{\theta \in \Theta} \int_{\Xi} \left\{ \int   \log\left(\dfrac{\ftrue(z | \xi )}{f(z | \theta )}\right) \ftrue(z | \xi )  \ dz \right\} \pi\left(\xi \left|z_{1:n}\right.\right) \ d\xi=\arg \max\limits_{\theta \in \Theta} \int \log f(z | \theta )  \ptrue(z | z_{1:n} ) \ dz.
\end{equation}
Exchanging differentiation and integration, we can deduce that the solution to \eqref{pp1} is also the solution to the estimating equation
\[
\int\frac{\partial \log f(z |t )}{\partial t}  \ptrue (z | z_{1:n} ) dz =\int S(z,t) \ptrue (z | z_{1:n} ) dz = 0
\]
where $S(z,\theta)$ is the score function. The minimization in \eqref{pp1} does not involve prior opinion concerning $\theta$, but \eqref{pp1} can be modified to
\[
\arg \max\limits_{\theta \in \Theta} \left\{ \int (\log f(z |\theta) + \log \pi_0 (\theta )) \ptrue (z | z_{1:n} ) \ dz \right\}
\]
or via the modified score function 
\[
S^*(z,\theta)=S(z,\theta)+ \frac{\partial }{\partial \theta}\log \pi_0\left(\theta\right).
\]
A sample from $\ptrue(z | z_{1:n} ) $ can be converted to a sampled value of $\theta$ in the same fashion as discussed in the main paper, which yields a fully Bayesian procedure with the solution of the usual likelihood-based posterior distribution.

\subsection{{Loss-based inference via the Gibbs posterior}}
\label{gibbs}
An inference mechanism can be derived directly using a loss function connecting the distribution of $Z$ and $\theta$ based on the definition in Equation (3) in the main paper.  The approach derives a probability measure of $\theta$ to define the posterior distribution $\pi \left(\theta \left|z_{1:n}\right.\right)$ given a prior $\pi_0(\theta)$.  This formulation \citep{zhang2006a,jiang2008gibbs,bissiri2016general,bissiri2019general} constructs posterior inference without the concept of likelihood, instead relying entirely on a loss specification, and the identification of a function, $\varphi$, that combines the aggregate loss across the observed data and the prior distribution such that $\pi \left(\theta \left|z_{1:n}\right.\right) =  \varphi \left\{\ell\left(z_{1:n},\theta\right),\pi_0(\theta)\right\}$.  \cite{zhang2006a} defined the objective function for loss-based inference with a probability measure $\mu$ on $\Theta$ as
\begin{align}
	\arg\inf\limits_{\mu \in \mathcal{M}_{\pi_0}} \left\{ \int_{\Theta} \ell \left(z,\theta\right)\mu\left(d \theta\right) + \right. & \frac{\mathcal{K}\left(\mu,\pi_0\right)}{\eta}  \left. \vphantom{\int} \right\} \nonumber \\[6pt]
	& =\arg\inf\limits_{\mu \in \mathcal{M}_{\pi_0}} \int \log \left[\frac{ \mu\left(\theta\right)}{\exp\left(-\eta\ell\left(z,\theta\right)\right)\pi_0\left(\theta\right)}\right]\mu\left(d \theta\right) \label{bobj}
\end{align}
where $\mathcal{M}_v$ is the space which is absolutely continuous with respect to $v$, $\ell$ is some measurable function, such that $ \ell(z,\cdot): \Theta \longrightarrow \mathbb{R}$ is measurable with respect to $\mu$ for every $z$ in the support, and $\mathcal{K}\left(\mu,\pi_0\right)$ is the KL divergence between two probability measures $\mu$ and $\pi_0$.   Parameter $\eta$, which has to be pre-specified,  controls the trade-off between the prior and the loss term.  The solution to the optimization problem in \eqref{bobj} is the so-called \textit{Gibbs posterior}
\begin{equation}
	\label{gibbssol}
	\pi  (\theta | z_{1:n} )= \frac{\exp \left(-\eta \ell \left(z_{1:n},\theta\right) \right) \times \pi_0\left(\theta\right)}{\displaystyle\int_{\Theta} \exp \left(-\eta\ell \left(z_{1:n},t\right) \right) \times \pi_0\left(dt\right)}
\end{equation}
defined if and only if the denominator is finite. The posterior distribution in \eqref{gibbssol} gives a formal Bayesian procedure to update prior beliefs on $\theta$ to posterior beliefs based on the loss function and decision-theoretic arguments.

The term $\exp \left(-\eta\ell \left(z_{1:n},\theta\right) \right)$ replaces the `likelihood' in conventional Bayesian updating.  This term does not necessarily correspond to a well-defined likelihood as it does not result from a probabilistic specification for the observable quantities.  In addition, to be considered a likelihood for conditionally independent observables, we must essentially assume that the parameter $\theta$ (rather than $\xi$) induces conditional independence, even though $\theta$ does not completely specify the data generating distribution.  Finally, unlike a true conditional probability model, the un-normalized quantity $\exp \left(-\eta\ell \left(z,\theta\right) \right)$ does not facilitate probabilistic prediction for future data, as it is not presumed integrable with respect to $z$.

\section{Asymptotics}
\subsection{Definition of Bayesian Consistency}
{\begin{definition} \citep{walker2001bayesian}
		For realizations $z_1, z_2,\ldots,z_n$ drawn independently from some unknown underlying distribution $\Ftrue$  with true data generating value $\thetatrue$ in the interior of the parameter space $\Theta$, the posterior mass assigned to $A \subseteq \Theta$ is given by
		\[
		\Pi^n\left(A\right)=\pi\left(\theta\in A \left|z_1,\ldots,z_n\right.\right)=\dfrac{\displaystyle \int_{A}R_n\left(\theta\right)\pi_0\left(d\theta\right)}{\displaystyle \int R_n\left(\theta\right)\pi_0\left(d\theta\right)}
		\]
		where
		\[
		R_n\left(\theta\right)=\prod_{i=1}^{n} \exp \left[- \left\{ \ell\left(z_i,\theta\right) - \ell\left(z_i,\theta^*\right) \right\} \right]
		\]
		and where $\pi_0\left(\theta\right)$ is the prior density for $\theta$. If $A_\epsilon=\left\{\theta: d(\theta,\theta^*)>\epsilon \right\}$ where $d(\theta,\thetatrue)$ is some distance measure, the posterior distribution is consistent in the Bayesian sense if $\Pi^n\left(A_\epsilon\right) \longrightarrow 0 $ almost surely under $\Ftrue$.
\end{definition}}

\subsection{Assumptions}

\begin{assumption}
	The loss function $\ell(\theta,z) :  \mathbb{Z}  \times \Theta \to \mathbb{R} $ is a measurable function, bounded from below, with 
	\[
	\int \ell\left(z,\theta\right) d\Ftrue\left(z\right) <\infty \qquad \forall \theta \in \Theta
	\]
	where $\Theta$ is a compact and convex subset of a $p$-dimensional Euclidean space.
\end{assumption}

\begin{assumption}
	$\ell\left(z,\theta\right)$ is continuous $\forall\theta \in \Theta$.
\end{assumption}

Let $\mathbf{U}_i\left(\theta\right)={\partial \ell\left(z_i,\theta\right)}\big/{\partial\theta ^{\top}} $.  Minimizing the expected loss function, $\E_{\Ftrue}[\ell\left(Z,\theta\right)]$, is equivalent to solving a $p \times 1$ system of estimating equations given by $\E_{\Ftrue}[\bU(\theta)] = \mathbf{0}$, with expectations taken with respect to the true data generating model $\Ftrue$.

\begin{assumption}
	$\thetatrue \in \Theta $ is the unique solution to $\E_{\Ftrue}[\bU(\theta)] = \mathbf{0}$, and for arbitrary $\delta >0$, there exists an $\epsilon>0$ so that
	\[
	\lim\limits_{n\longrightarrow \infty} \mathbb{P}\left(\sup_{\norm{ \theta  - \thetatrue} \ > \ \delta}\left|\frac{1}{n}\sum_{i=1}^{n}\left(\ell(Z_i,\theta)-\ell(Z_i,\thetatrue)\right)\right| < \epsilon\right) =1
	\]
	where $Z_1,\ldots,Z_n \sim \Ftrue$.
\end{assumption}

\begin{assumption}
	$\E\left[\sup_{\theta\in \Theta}\norm{\mathbf{U}\left(\theta\right)}^{\gamma}\right] <\infty$ for $\gamma>2$.	Suppose there exists a neighborhood, $\widetilde{\Theta}$ of $\thetatrue$ within which ${\bU}(\theta)$ is continuously differentiable.
	\[
	\E_{\Ftrue}\left[\sup_{\theta \in \Theta^*}\norm{\dot{\bU}(\theta)}_{F}\right]< \infty,
	\]
	with $\norm{\cdot}_{F}$ denoting the Frobenius norm.
\end{assumption}

\begin{assumption}
	There is an open ball $B$ containing $\thetatrue$ such that all first, second and third partial derivatives of $\ell(\theta,z)$ with respect to $\theta \in B$ exist and are continuous for all $z$. Furthermore, there exist measurable functions $G_j$, $G_{jk}$, $G_{jkl}$ and $M_{jkl}$ such that for $\theta \in B$ we have
	\begin{align*}
		\left|\frac{\partial \ell\left(z,\theta\right)}{\partial \theta_j }\right| &\leq G_j(z) & \text{with } \displaystyle \int G_j(z) d\Ftrue\left(z\right) <\infty, \\[6pt]
		\left|\frac{\partial^2 \ell\left(z,\theta\right)}{\partial \theta_j \partial \theta_k}\right| &\leq G_{jk}(z) & \text{with } \displaystyle \int G_{jk}(z) d\Ftrue\left(z\right) <\infty, \\[6pt]
		\left|\frac{\partial^3 \ell\left(z,\theta\right)}{\partial \theta_j \partial \theta_k\partial \theta_l}\right| & \leq G_{jkl}(z) & \text{with } \displaystyle \int G_{jkl}(z) d\Ftrue\left(z\right) <\infty, \\[6pt]
		\left|\frac{\partial \ell\left(z,\theta\right)}{\partial \theta_j }\frac{\partial^2 \ell\left(z,\theta\right)}{\partial \theta_k \partial \theta_l}\right| &\leq M_{jkl}(z) & \text{with } \displaystyle \int M_{jkl}(z) d\Ftrue\left(z\right) <\infty.
	\end{align*}
\end{assumption}	
Let
\[
\mathcal{I}(\theta^\ast) = \E[ \bU(\theta^\ast) \bU(\theta^\ast)^\top] \qquad \qquad \mathcal{J}(\theta^\ast) = -\E[\dot{\bU}(\theta^\ast)]
\]
both $(p \times p)$ matrices, and $\dot{\bU}(\theta^\ast) = {\partial \bU(\theta)}\big/{\partial \theta^\top} \big|_{\theta = \theta^\ast}$

\begin{assumption}
	$\mathcal{I}(\theta) $ and $\mathcal{J}(\theta)$ are non-singular and $\mathcal{J}$ is full rank, rank$\left(\mathcal{J}(\theta)\right)=p$, for $\theta \in B$, with all elements finite.
\end{assumption}

\subsection{Proof of Theorem 1}
\begin{proof}
	Suppose that $\theta^s_1$ is the minimizer of the weighted loss
	\begin{equation*}
		\theta^s_1 \equiv \theta (\varpi^s,\zeta^s) =  \arg\min_{t}\sum_{k=1}^{\infty}\varpi_k^s\ell\left(\zeta_k^s,t\right).
	\end{equation*}
	given by the prior-to-posterior computation.  From Theorem 1 in \cite{lijoi2004extending}, there exists an unique random element $F_1$ such that
	\[
	\sum_{k=1}^{\infty}\varpi_k^s \ell (\zeta_k^s, \theta^s )  \longrightarrow \min\limits_{\theta \in \Theta}\int \ell\left(z,t \right) dF_1\left(z\right)\;\;\; n \longrightarrow \infty.
	\]
	
	It remains to show that $F_1 \equiv \Ftrue$. As $F_1$ is a draw from the posterior distribution under a Bayesian non- parametric formulation given by the prior-to-posterior computation,  according to the de Finetti's representation theorem, the posterior distribution will be degenerate at $\Ftrue$ as $n \longrightarrow  \infty$.  Therefore, $\theta(\varpi^s)$,  is unique for any given $\zeta^s$ and $\varpi^s$, $\theta(\varpi^s)$ will be become degenerate at $\thetatrue$ as $n\longrightarrow \infty$.
	
	Alternatively, suppose that $\theta^s_2$ is the minimizer of the Monte Carlo estimate of the posterior predictive expectation
	\begin{equation*}
		\theta^s_2 \equiv \theta\left(z^s\right) =\arg\min_{t}  \sum_{i=1}^{N}\ell\left(z^{s}_i,t\right).
	\end{equation*}
	Again by Theorem 1 in \cite{lijoi2004extending}, there exists an unique random element $F_2$ such that
	\[
	\sum_{k=1}^{N} \ell (z_k^s, \theta^s_2 )  \longrightarrow \min\limits_{\theta \in \Theta}\int \ell\left(z,\theta \right) dF_2\left(z\right)\;\;\; n,N \longrightarrow \infty.
	\]
	These results confirm that the posterior distribution generated by each approach converges weakly to a probability measure with all its mass on $F_2 \equiv p(z^s_{1:N}| z_{1:n} )$, and it also remains to show that $F_2 \equiv \Ftrue$. Under mild regularity conditions and the correct specification of the model leading to $\pi ^\ast(\xi \mid z_{1:n})$, with $\widehat\xi\left(z_{1:n}\right)$ in a neighbourhood of $\xi^\ast$, following \citet{bernardo1979reference}, we have
	\begin{equation*}
		\begin{aligned}
			\ptrue(z^s_{1},\ldots,z^s_{N} | z_{1:n} ) &= \ptrue (z^s_{1},\ldots,z^s_{N} \mid \hat \xi (z_{1:n} ) ) + \textrm{O}(1)\\
			&= \ptrue(z^s_{N}  \mid z^s_{1},\ldots,z^s_{N-1},\xistar)\cdots   \ptrue( z^s_{1}  \mid \xistar )+ \textrm{o}(1) &  n\longrightarrow \infty\\
			&=\ftrue(z^s_{N}  \mid \xistar)\cdots \ftrue( z^s_{1}| \xistar)+ \textrm{o}(1)
		\end{aligned}
	\end{equation*}
	and therefore,  a draw from the predictive $p(z^s_{1},\ldots,z^s_{N} | z_{1:n} )$ suitably simulates a collection of $N$ sample points from the true data generating model $\Ftrue(z|\thetastar)$ as $n\longrightarrow \infty$.  Therefore, $F_2$ is the same as $\Ftrue$.   As the solution, $\hat\theta(z^s)$,  is unique for any given $z^s$, therefore,  $\hat\theta(z^s)$ will become degenerate at $\thetatrue$ as both $N\longrightarrow \infty$ and $n\longrightarrow \infty$.
\end{proof}

\subsection{Proof of Theorem 2}
\begin{proof}
	First, we define $S_N(\theta)= \sum_{i=1}^N{\partial \varpi_i \ell\left(z_i^s,\theta\right)}\big/{\partial\theta ^{\top}} = \sum_{i=1}^N \varpi_i\bU_i(\theta)$, and $\mathcal{J}_{\varpi} (\theta)= -\sum_{i=1}^N \varpi_i\dot{\bU_i}(\theta)$. Then the Taylor expansion of $S_N(\hat \theta_n)$ around $\hat\theta(\varpi)$ becomes
	\begin{equation}
		\label{score}
		S_N(\hat \theta_n) = (\mathcal{J}_{\varpi} (\hat \theta_n) - R_n) ( \theta^s - \hat \theta_n)
	\end{equation}
	where $R_n$ is the reminding term.  For a large $n$,  $R_n$ is negligible under regularity conditions. To see what remains to be proved, we rewrite \eqref{score} as
	\[
	\vartheta_{n,N}^s = 	\sqrt{N}	( \theta^s - \hat \theta_n)  = (\mathcal{J}_{\varpi} (\hat \theta_n) - R_n) ^{-1}\sqrt{N} S_N(\hat \theta_n).
	\]
	By Lemma 7 in \cite{newton1991weighted},
	\[
	\mathcal{J}_{\varpi} (\hat \theta_n)   \xrightarrow{p}  \mathcal{J} (\thetatrue).
	\]
	For any $t \in \mathbb{R}^{p}$ with $\left|t\right| =1$, we defined  $m_N(t) = \sqrt{N}t^{\top}S_N(\hat \theta_n) $, and by Theorem 2 in \cite{ishwaran2002exact}, which approximates the DP using a finite dimensional Dirichlet process,
	\begin{equation*}
		\begin{aligned}
			m_N(t) &\approx \sqrt{N} \sum_{j=1}^{p} t_j \left(\frac{\sum_{i=1}^{N} H_i U_i(\hat \theta_n) }{\sum_{i=1}^{N} H_i }\right) \\
			&= \frac{1}{\bar H_N} \frac{\sum_{i=1}^{N} a_{in}H_i }{\sqrt{N}}
		\end{aligned}
	\end{equation*}
	where $H_i$ is iid Exponential$\left(\alpha/N\right)$ random variables independent of the data $z^s$,  $\bar H_N =1/N\sum_{i=1}^{N}H_i$, and $a_{in}= \sum_{j=1}^{p}t_jU_i(\hat \theta_n)$. By Lindeberg-Feller central limit theorem and Lemma 8 in \cite{newton1991weighted}, we have
	\[
	\frac{\sum_{i=1}^{N} a_{in}H_i }{\sqrt{N}}  \xrightarrow{d} Normal_p(\mathbf{0},(\frac{\alpha}{N})^2t^{\top} \mathcal{I}(\thetatrue) t).
	\]
	By  Slutsky’s theorem, with $\bar H_N \to {\alpha}/{N}$, we have
	\[
	m_N(t)  \xrightarrow{d} Normal_p(\mathbf{0},t^{\top} \mathcal{I}(\thetatrue) t)  \text{    as    } n,N\to \infty.
	\]
	Therefore, by Cramer-Wold theorem, we have
	\[
	\sqrt{N} S_N(\hat \theta_n)   \xrightarrow{d} Normal_p(\mathbf{0},\mathcal{I}(\thetatrue))   \text{    as    } n,N\to \infty.
	\]
	By  applying Slutsky’s theorem again, we have
	\[
	\vartheta_{n,N}^s   \xrightarrow{d} Normal_p(\mathbf{0},\mathcal{J} (\thetatrue)^{-1}\mathcal{I}(\thetatrue)\mathcal{J} (\thetatrue)^{-\top})   \text{    as    } n,N\to \infty.
	\]
\end{proof}	

\subsection{Proof of Corollary 1}
\begin{proof}
	When we have a mis-specified PS model and a correctly specified OR model, $\theta^\ast=\left(\psi^\ast,\beta^\ast,0\right)$. The results follow by applying Theorem 2.  When the outcome model is mis-specified but the PS model is correctly specified, $X \perp \!\!\! \perp  D \left| e\left(x; \gamma^\ast\right) \right.$ and $\hat \gamma \longrightarrow \gamma^\ast$. Therefore, $e\left(x;\hat \gamma\right)$ is an asymptotic balancing score. Suppose we specify the mean model as above and assume that the effect of $D$ is captured via the term $\psi D$.  Under the assumption of no unmeasured confounding, we can find  $\theta^\ast=\left(\psi^\ast,\beta^\ast,\phi^\ast\right)$ under the specified loss function corresponding to such mis-specified OR models.  This is again in line with the standard frequentist approach for mis-specified models. Therefore, we can construct the same asymptotic results as for the mis-specified PS case by applying Theorem 2.
\end{proof}

\subsection{P\'{o}lya urn scheme representation}

\cite{fong2021martingale} stated two conditions which the predictive distribution has to satisfy.

\smallskip
{\begin{condition}
		The sequence of predictive distributions, $p_{n+1}(y|d,x),p_{n+2}(y|d,x),\ldots$, converges almost surely to a random probability distribution $p_{\infty}(y|d,x)$, for all $y\in \mathbb{R}$.
	\end{condition}
}
\smallskip
{\begin{condition}
		The posterior expectation of the random $p_{\infty}(y|d,x)$ satisfies $\E\left[p_{\infty}(y|d,x)\left|z_{1:n} \right.\right]=p_{n}(y|d,x)$
		almost surely for all $y\in \mathbb{R}$.
	\end{condition}
}
\smallskip

{Assuming these two conditions, $p_{\infty}(y|d,x)$ is considered as the best estimate of the unknown true data generating mechanism under the specified model sequence, and gives a mechanism for generating posterior uncertainty of $\theta$ without applying Bayes rule. \cite{berti2006almost} showed that the conditional distribution, $p_{n+N}(y|d,x)$, converges weakly to a random probability measure almost surely for each pair of $(d,x)$ if these two conditions are satisfied.
}

{In the predictive resampling approach derived from the Dirichlet process and indicated in Equation (7) in the main paper, the sequence $\{G_j\}_{j=1}^N$ are precisely predictive models that align with the theory of \cite{berti2006almost}, and therefore we have the following theorem.
}\smallskip
{\begin{theorem}
		There exists a random probability measure $G_{\infty}$ such that $G_{n+N}$ converges weakly to $G_{\infty}$.
	\end{theorem}
}
%\smallskip
\begin{proof}
	For the sequence of random probability measures based on the DP construction $\left\{G_{N},G_{N+1},\ldots\right\}$ defined on the probability space $(\Omega,A,\P)$, take values in the measurable space $(\mathbb{Y},\mathcal{Y})$, we define
	\[
	G_{N}\left(f\left|x,d\right.\right)=\int f(y) d G_{N}\left(y\left|x,d\right.\right)\;\; \text{all bounded measurable }f:\mathbb{Y}\to  \mathbb{R}.
	\]
	This integral is finite if $\int \log(1+\left|f(y)\right|) d G_{0}\left(y\left|x,d\right.\right)<+\infty$   \citep{feigin1989linear}. We denote a filtration, $\mathcal{F}_{i} = \sigma(Z_1,\ldots,Z_i)$. Taking the conditional expectation, from Fubini’s theorem, we have
	\[
	\E\left[G_{N+1}\left(f\left|x,d\right.\right)\left|\mathcal{F}_{N}\right.\right]= \int f(y) \E\left[dG_{N+1}\left(y\left|x,d\right.\right)\left|\mathcal{F}_{N}\right.\right] = G_{N}\left(f\left|x,d\right.\right)
	\]
	because $G_{N}\left(y\left|x,d\right.\right)$ is a martingale with respect to $\mathcal{F}_{N}$ regardless of the draw for the pair of $x,d$. As $f$ is bounded, then $ \E\left[\left|G_{N}\left(f\left|x,d\right.\right)\right|\right]$  is also bounded. Therefore,  $G_{N}\left(f\left|x,d\right.\right)$ is also a martingale  with respect to $\mathcal{F}_{N}$. By Theorem 2.2 in \cite{berti2006almost}, so there exists a random probability measure $G_{\infty}$ defined on $(\Omega,A,\P)$ such that $G_{N} \to G_{\infty}$ weakly almost surely.
\end{proof}
This theorem confirms that predictive resampling via the Dirichlet process is a valid Bayesian update and gives the same uncertainty quantification as the prior-to-posterior update.  From Equation (5) in the main paper, we may deduce that the value obtained from solving the minimization problem in Equation (6) in the main paper is a sample from the posterior distribution of the target parameter.

\section{Additional simulation results}
{In this example, we examine the performance of the proposed updating approaches under some extreme PS distributions, with binary exposure, but where there is no treatment effect.}  The data are simulated as follows: we simulate $X_1,X_2 \sim \mathcal{N}\left(1,1\right)$ and $X_3,X_4 \sim \mathcal{N}\left(-1,1\right)$ independently, and then simulate
\begin{equation*}
	\begin{aligned}
		D\left|X_1,X_2,X_3,X_4 \right. &\sim \text{ Bernoulli}\left(\text{expit}\left(\gamma_0+\gamma_1X_1+\gamma_2X_2+\gamma_3X_3+\gamma_4X_4\right)\right)\\
		Y \left|D,X_1,X_2,X_3,X_4\right. &\sim  \mathcal{N}\left(0.25X_1+0.25X_2+0.25X_3+0.25X_4+1.5X_3X_4,1\right).\\
	\end{aligned}
\end{equation*}
In the analyses, the PS model is assumed to be correctly specified. For the {outcome model}, we fit the model with the treatment indicator and estimated PSs only as covariates. To investigate how the PS distribution affects the estimation of the treatment effect, different PS distributions are considered:
\begin{itemize}
	\item Scenario A: $\gamma=\left(0.00,0.30,0.80,0.30,0.80\right)$, generating a nearly uniform distribution of propensity scores;
	\item Scenario B: $\gamma=\left(0.50,0.50,0.75,1.00,1.00\right)$, having a greater density of lower scores;
	\item Scenario C: $\gamma=\left(0.00,0.45,0.90,1.35,1.80\right)$, having very few high scores.
\end{itemize}
In this example, we also fit Bayesian regression on the correctly specified OR.  

Table~\ref{sim2ss} summarizes the mean estimates of $\theta$ over 1,000 Monte Carlo replicates for three different scenarios described above. For a correctly specified OR, the coverage rate is around the nominal level. For the proposed methods, the results suggest that all approaches yield unbiased estimates across all scenarios (see Appendix D).  As in Example 1, under a fairly uniform PS distribution, these approaches indicate nearly the same performance, and the results agree in terms of posterior mean and variance. However, when the PS distribution is slightly skewed (Scenario B), Method II exhibits a slightly higher bias and greater variance, notably when $n$ is small but these differences diminish as $n$ increases.  The bias and greater variance in Method II become more obvious when the PS distribution is highly skewed (Scenario C). Also in Scenario C, Method I has consistently the smallest variance. In general, Methods II and III have very similar performance in those scenarios.

\begin{table}
	\caption{\label{sim2ss} Simulation results of the marginal causal contrast, with true value equal to 0, on 1000 simulation runs on generated datasets of size $n$.  Bayes-OR represents standard Bayesian inference for the correctly specified OR with non-informative priors.}
	\centering
	\begin{tabular*}{40pc}{@{\hskip5pt}@{\extracolsep{\fill}}l@{}c@{}c@{}c@{}c@{}|c@{}c@{}c@{}c@{}|c@{}c@{}c@{}c@{\hskip5pt}}
		\hline
		& \multicolumn{4}{c}{Scenario A}&\multicolumn{4}{c}{Scenario B}&\multicolumn{4}{c}{Scenario C} \\ \cline{2-13}
		$n$	&100 &200  & 500 &1000	&100 &200 & 500 &1000	&100 &200 & 500 &1000\\
		\hline	
		\multicolumn{13}{l}{Mean}	\\
		\hline	
		Method I&-0.004 &-0.005 &-0.010 &0.002 & 0.001 &-0.007&0.005& 0.003& -0.007&-0.006&0.004&0.000\\
		Method II& -0.010  &-0.001&0.003&-0.002 &-0.001 &0.007 &-0.002  &-0.002&0.038&0.024& 0.000&0.002\\
		Method III &-0.003 & -0.004&0.001& 0.004&-0.007&0.006& 0.000&-0.008& 0.054&-0.008&0.012& 0.002\\
		Bayes-OR & 0.000 & -0.001 & 0.002 & -0.004 &  0.008& 0.001&-0.004 & -0.007 & -0.005 & 0.002 &0.000 & 0.002\\
		\hline
		\multicolumn{13}{l}{Variance}	\\
		\hline	
		Method I &0.148 &0.067&0.028&0.013&0.154&0.079&0.031&0.015&0.144&0.069&0.043&0.014\\
		%Method II & 0.156 &0.076&0.027&0.015&0.165 &0.077&0.033&0.016&0.219&0.107&0.046&0.022\\
		Method II& 0.155& 0.071&0.030 & 0.015& 0.163 &0.080&0.032& 0.016& 0.233&0.112&0.043&0.023\\
		Method III &  0.145&  0.074&0.029&0.014 &0.139 &  0.080 & 0.032&0.015& 0.234& 0.109&0.044&0.021\\
		Bayes-OR & 0.055 & 0.026 & 0.010 & 0.006 &  0.066& 0.031&0.012 & 0.006 & 0.087 & 0.040 &0.017 & 0.008\\
		\hline
		\multicolumn{13}{l}{Coverage, \%}	\\
		\hline	
		Method I&94.4 & 95.2 &94.8 & 94.4 & 93.8&  95.0&95.2&94.5& 96.1& 94.0&95.2 &  94.7\\
		Method II& 91.6&  93.1&94.4 &94.7 &91.5 & 93.0 &94.8 & 95.1 &  89.2&92.8&94.7& 94.3\\
		Method III & 93.6  & 95.9 & 95.5& 95.8 &95.1& 94.9 & 95.9 &95.5 &91.0 & 94.6 &94.5& 95.1\\
		Bayes-OR & 95.2 & 96.0 & 95.9 & 94.4 & 94.9& 94.3 & 95.2 & 94.0 & 94.2 & 96.0 & 94.2 & 95.0\\
		\hline
	\end{tabular*}
\end{table}

\section{Application: Posterior predictive densities for different methods}
\begin{figure}[ht]
	\centering
	\includegraphics[scale=0.5]{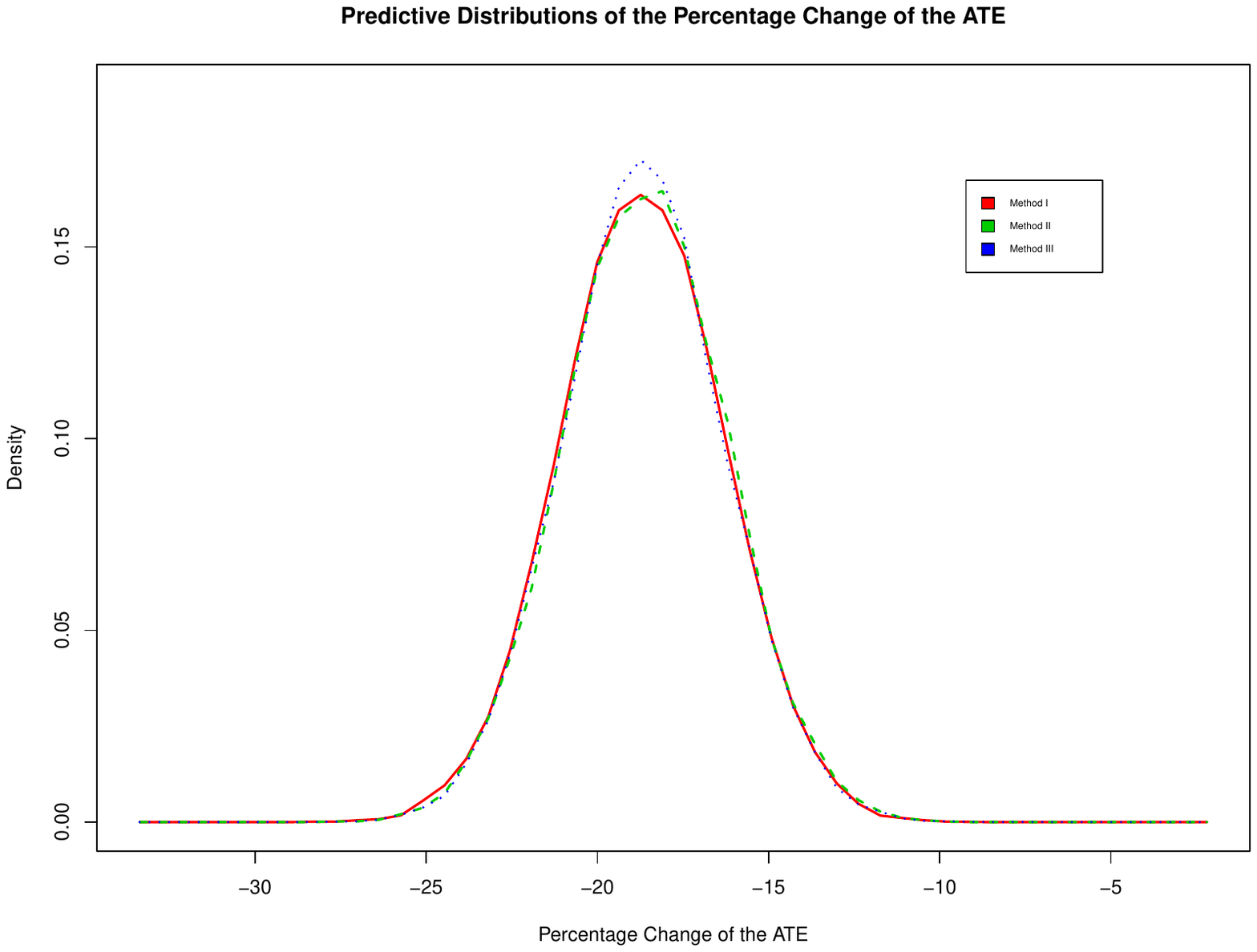}
	\caption{\label{ateapp}UK speed camera data: Posterior predictive densities using three different methods. }
\end{figure}

\end{document}